\documentclass[twocolumn,showpacs,amsmath,amssymb,superscriptaddress]{revtex4}

\usepackage{graphicx} 
\usepackage{dcolumn}  
\usepackage{bm}       

\newenvironment{new}
{\begin{itemize}
\setlength{\itemsep}{1pt}}
{\end{itemize}}

\def\be{\begin{eqnarray}}
\def\ee{\end{eqnarray}}
\def\bse{\begin{subequations}}
\def\ese{\end{subequations}}
\def\bi{\begin{new}}
\def\ei{\end{new}}
\def\l{\langle}
\def\r{\rangle}
\def\a{\alpha}
\def\b{\beta}
\def\p{\prime}
\def\ap{\alpha^{\prime}}
\def\bp{\beta^{\prime}}

\def\ra{\longrightarrow}
\def\bs{\boldsymbol}
\def\E{{\cal E}}

\newcommand{\ket}[1]{\left| {#1} \right>}
\newcommand{\bra}[1]{\left< {#1} \right|}

\begin{document}

\title{Realistic fast quantum gates with hot trapped ions}

\author{Marek \v{S}a\v{s}ura}
\affiliation{Centre for Quantum Computation, Clarendon Laboratory,
Department of Physics, University of Oxford, 
Parks Road, Oxford OX1 3PU, UK}
\affiliation{Research Center for Quantum Information (RCQI),
Institute of Physics, Slovak Academy of Sciences,
D\'{u}bravsk\'{a} cesta 9, Bratislava 842 28, Slovakia}

\author{Andrew M. Steane}
\affiliation{Centre for Quantum Computation, Clarendon Laboratory,
Department of Physics, University of Oxford, 
Parks Road, Oxford OX1 3PU, UK}

\date{December 2, 2002}

\begin{abstract}
The ``pushing gate'' proposed by Cirac and Zoller for quantum logic
in ion traps is discussed, in~which a force is used to give 
a~controlled push to a pair of trapped ions and thus realize a~phase
gate. The original proposal had a weakness 
in that it involved a hidden extreme sensitivity to the~size of 
the~force. Also, the physical origin of this force was not fully addressed.
Here, we~discuss the sensitivity and present a~way
to avoid it by choosing the spatial form of the pushing force
in an optimal way. We also analyse the effect of
imperfections in a~pair of $\pi$~pulses which are used to implement
a ``spin-echo'' to cancel correlated errors.
We present a physical model for the force, namely
the dipole force, and discuss the impact of unwanted photon
scattering, and of finite temperature of the ions. The main effect
of the temperature is to blur the phase of the gate owing to the~ions
exploring a range of values of the force. When the distance scale
of the force profile is smaller than the ion separation
this effect is more important than the high-order terms in
the~Coulomb repulsion which were originally discussed.
Overall, we find that whereas the ``pushing gate'' is not as
resistant to imperfection as was supposed, it remains 
a~significant candidate for ion trap quantum computing since it does
not require ground state cooling, and in some cases it does not
require the Lamb-Dicke limit, while the gate rate is fast, close to
(rather than small compared~to) the trap vibrational frequency.
\end{abstract}


\maketitle

\section{Introduction}

Since the first recognition of the advantages of ion traps for
quantum computing \cite{CZ95}, various proposals to achieve quantum
gates between pairs of trapped ions have been put forward~\cite{exp}.
In all cases each ion stores one or more qubits in its internal state,
and most proposals have envisaged two or more ions in the same
harmonic well, with the joint motional degree of freedom (normal
mode of oscillation) serving as a further carrier of quantum
information, which can be coupled to any chosen ion by laser
excitation~\cite{steane97,wineland98, nagerl00,tut}.
More recently a~method of a qualitatively different
form was proposed, in which the ions need not be in the same
harmonic well, and a quantum gate is achieved through a more
direct use of the Coulomb repulsion between ions \cite{ions1}.
Two ions in neighbouring harmonic wells are pushed by a force
which depends on their internal state.
Therefore, their mean separation depends on
their internal state while the force acts. It is shown in
Ref.~\cite{ions2} that the phase acquired from the Coulomb energy has the
form of single-qubit rotations combined with the controlled-phase
gate $\ket{00}\bra{00}+\ket{01}\bra{01}+\ket{10}\bra{10}+
e^{i\vartheta}\ket{11}\bra{11}$. 
This ``pushing gate'' is significant for three
reasons. First, it was shown~\cite{ions2} that in the right
conditions the gate is unusually insensitive to the details of the
motion of the~ions, i.e. the quantum phases depend primarily
on the~mean positions of the ions, and are insensitive, for example, to
the ion temperature. Secondly, the gate can be faster than those
previously proposed.
Thirdly, the fact that the gate can act on
ions in separate wells is useful for scaling the computer up to
large numbers of ions, because it is possible to move the ions
around without ever having to ``tease apart'' two ions which were
initially in the same well.

For these reasons the ``pushing gate'' is a promising idea, but
various aspects were left unclear in work up till now, as follows.

(i) The method generates single-qubit rotations simultaneously with
the desired two-qubit phase gate. In the~original work \cite{ions2}
it was assumed that these rotations would be unproblematic, since they
are known from the~experimental parameters and can be undone
afterwards, if so desired, by any available single-qubit gate
method. Therefore, the explicit calculation of these phases was
not carried out. However, if the rotation angles are large
compared to $\pi$ they will be sensitive to experimental
imprecision. Therefore, it is necessary to calculate them and
if possible, find ways to suppress or compensate for them.

(ii) A state-dependent pushing force was introduced without
analysing in sufficient detail the physical mechanism which gives
rise to the force.

(iii) The proposal assumed that the force is independent of position,
and it was unclear to what extent the~good behaviour of the gate relied
on that assumption.

(iv) The effect of imprecision or fluctuations in the~parameters was
not considered.

The reason it is important to clarify these points is that until
this is done, it is not clear whether the good performance of
the gate is genuine, or whether it is an~artifact of some
unphysical assumption, such as an unreasonably high laser
intensity or an unreasonably high precision in parameters.
This paper will address the points just listed.

In Sec.~\ref{gc}
we introduce the basic concept of a generalised phase
gate in order to bring out the physics without reference to any
specific physical system. We also examine the double $\pi$-pulse
``spin-echo'' which was suggested in Ref.~\cite{ions2} as a way to
improve the fidelity. The ``spin-echo'' is used to cancel
a~correlated error which might be large, so there is a danger that
a~small imprecision in this part of the evolution might result in
a~large loss of fidelity. Therefore, we examine the influence of
imprecision in the~$\pi$~pulses.

In Sec.~\ref{ions}
we introduce the ion trap system. First, we show how the gate works in
a qualitative way, and then we write down the Hamiltonian and
deduce the~classical trajectories of the ions under
the influence of the~two trapping potentials, their Coulomb repulsion,
and the~state-selective and time-dependent pushing force which is, 
for this section, assumed to be independent of position.

Sec.~\ref{dp} then obtains all the quantum phases which arise
in the evolution in the semiclassical approximation,
where the trajectories are
classical and the quantum phase is obtained from a path integral
along the~classical trajectory. Our main purpose is to obtain
the~single-qubit rotation phases which were not calculated in
the~original paper \cite{ions2}. In order to make
the treatment self-contained we
also rederive the two-qubit phase which is the central feature of
the gate.

Sec.~\ref{force} considers an explicit choice for the
pushing force. The most natural choice is the dipole force
from a~non-resonant laser beam. This has the unavoidable
consequence that unwanted scattering of photons will occur, which
reduces the fidelity. We calculate the photon scattering and hence
the loss of fidelity.

In Sec.~\ref{fluct} we examine the influence of
fluctuations in the laser intensity. This is important because
the single-qubit phases introduced by the gate are large, so that in
some parameter regimes small changes in laser intensity can cause
single-qubit rotations much larger than~$\pi$,
destroying the fidelity.
We propose a way to largely circumvent this problem by appropriately
choosing the~directions of the forces and the distance scale
of the potential energy function (e.g. AC Stark light shifts), which provides
the force.

In Sec.~\ref{non-uni} we examine the effect of
position dependence of the force, which is a major constraint on 
the~temperature of the ions. It is also non-negligible since
tightly focused laser beams, or a laser standing wave, must be
used to get sufficient dipole force without unreasonably high
laser power.

In Sec.~\ref{infid} we consider the total fidelity of the gate,
including the effects of imperfect~$\pi$~pulses, photon
scattering, positional dependence of the force, and
non-zero temperature of the ions. We give example values for
the calcium ion, and discuss the results in comparison to other
gate methods in ion traps.

\section{General concept}
\label{gc}

In this section we shall introduce a general concept of
a~two-qubit phase gate. We discuss the fidelity of the~phase gate
and also the fidelity of the gate incorporating a double $\pi$-pulse
or ``spin-echo'' method to cancel one source of infidelity.
Finally, we treat the latter case with imperfect $\pi$ pulses.

\subsection{Phase gate}

Consider a system of two interacting qubits described by the evolution
operator $G$ in the computational basis $\{|00\r, |01\r, |10\r, |11\r\}$
having the form
\be
\label{gc1}
G
&=&
\sum_{\a,\b=0}^1
|\a\b\r\l\a\b |\, e^{i\Theta_{\a\b}}\nonumber\\[1mm]
&=&
{\rm diag}\{
e^{i\Theta_{00}},
e^{i\Theta_{01}},
e^{i\Theta_{10}},
e^{i\Theta_{11}}
\}\,,
\ee
where the phases $\Theta_{\a\b}$ have
different values in general because the states
$|\a\b\r\equiv |\a\r_1\otimes |\b\r_2$
can have a different energy during the interaction and hence a different
action integral. A physical system which satisfies the evolution~(\ref{gc1})
will be introduced and described in Sec.~\ref{ions}.

The operation $G$ can be transformed into a two-qubit phase gate
$P_{\vartheta}$ which generates a phase $\vartheta$ if and only if
both qubits are in the logical state $|1\r$, that is
\be
\label{gc2}
P_{\vartheta}
&=&
{\rm diag}\{
1, 1, 1, e^{i\vartheta}
\}\,.
\ee
The transformation $G\rightarrow P_{\vartheta}$ can
be accomplished by local operations (single-qubit rotations)
applied on each qubit
\be
\label{gc4}
S=S_1\otimes S_2\,,
\ee
with
\bse
\label{gc5}
\be
S_1 &=& Z_1(\Theta_{10} - \Theta_{00} )\,,\\
S_2 &=& Z_2(\Theta_{01} - \Theta_{00})\,,
\ee
\ese
where
\be
\label{Zrot}
Z_j(\Theta)\equiv
|0\r_j\l 0|\,e^{i\Theta/2}+ |1\r_j\l 1|\,e^{-i\Theta/2}
\ee
is a rotation by the angle $\Theta$ about the $z$ axis of the Bloch sphere.
Bringing together
Eq.~(1) and (\ref{gc4}) we can realise the~phase
gate~(\ref{gc2}) in the following way
\be
\label{gc7}
\begin{array}[c]{lclcr}
|00\r & \stackrel{G}{\ \ra\ } &
e^{i\Theta_{00}}\,|00\r &
\stackrel{S}{\ \longrightarrow\ } & |00\r\,,\\[1mm]
|01\r & {\ra} & e^{i\Theta_{01}}\,|01\r &
{\ra} & |01\r\,,\\[1mm]
|10\r & {\ra} & e^{i\Theta_{10}}\,|10\r &
{\ra} & |10\r\,,\\[1mm]
|11\r & {\ra} & e^{i\Theta_{11}}\,|11\r &
{\ra} & e^{i\vartheta}|11\r\,,
\end{array}
\ee
where
\be
\label{gc8}
\vartheta=\Theta_{11}-\Theta_{10}-\Theta_{01}+\Theta_{00}\,,
\ee
and a global phase $(\Theta_{01} + \Theta_{10})/2$ was neglected.
The~two-qubit phase gate with
the~phase $\vartheta=\pi$ is equivalent (up to additional single-qubit rotations)
to a two-qubit controlled-{\sf NOT} ({\sf CNOT}) gate, which together with
single-qubit rotations forms a universal set of gates for quantum
computation \cite{barenco}.

\subsection{Imprecision of the phase gate}
\label{impr}

In practice for a particular physical system the phases $\Theta_{\a\b}$ in
Eq.~(1) are not accurately known and they may vary from one
realisation of the gate to another. On the~other hand, we can assume the
single-qubit rotations~(\ref{gc5}) are applied with
a very high accuracy. A good choice for the phases $\Theta_{\a\b}$
in Eq.~(\ref{gc5}) is obtained by estimating the most likely values
of these phases which will occur during the gate operation~(\ref{gc1}).
We denote this choice in the following way
\be
\label{gc9}
\begin{array}{ll}
\Theta_{00}\ \rightarrow\ \bar{\Theta}_{00}\,, &\qquad
\Theta_{01}\ \rightarrow\ \bar{\Theta}_{01}\,,\\[1mm]
\Theta_{10}\ \rightarrow\ \bar{\Theta}_{10}\,, &\qquad
\Theta_{11}\ \rightarrow\ \bar{\Theta}_{11}\,.
\end{array}
\ee
Then, the actual gate is not a perfect gate but it is rather the~transformation
\be
\label{gc10}
\begin{array}{lcrcr}
|00\r & \stackrel{G}{\ \ra\ } &
e^{i\Theta_{00}}\,|00\r &
\stackrel{S}{\ \ra\ } &
e^{i(\Theta_{00}-\bar{\Theta}_{00})}\,|00\r\,,\\[1mm]
|01\r & \ra &
e^{i\Theta_{01}}\,|01\r &
\ra &
e^{i(\Theta_{01}-\bar{\Theta}_{01})}\,|01\r\,,\\[1mm]
|10\r & \ra &
e^{i\Theta_{10}}\,|10\r &
\ra &
e^{i(\Theta_{10}-\bar{\Theta}_{10})}\,|10\r\,,\\[1mm]
|11\r & \ra &
e^{i\Theta_{11}}\,|11\r &
\ra &
e^{i\bar{\vartheta}}\,e^{i(\Theta_{11}-\bar{\Theta}_{11})}\,|11\r\,,
\end{array}
\ee
where we denoted
\be
\label{gc11}
\bar{\vartheta}=
\bar{\Theta}_{11}-\bar{\Theta}_{10}-\bar{\Theta}_{01}+\bar{\Theta}_{00}\,,
\ee
and we will require $\bar{\vartheta}=\pi$.
The precision of the gate is limited by the degree to which $G$ is not
perfectly predictable from one realisation of the gate to another,
owing to technical noise and the finite temperature of the~trapped ions.

\subsection{Fidelity of the phase gate}

In order to calculate the fidelity of the phase gate we have to compare
an~output state of a~perfect gate and an~actual gate.
We will define the fidelity as follows
\be
\label{gc12}
{\cal F}=
\left\l
\min|\l\Psi_{\rm perf}|\Psi_{\rm act}\r|^2
\right\r,
\ee
where the perfect state
is produced by the perfect phase gate~(\ref{gc2}) for $\vartheta=\pi$,
giving
\be
\label{gc-perf}
|\Psi_{\rm perf}\r = P_{\pi}|\Psi_0\r\,,
\ee
while the actual state follows from the actual gate given by
the~transformation~(\ref{gc10}), that is
\be
\label{gc-act}
|\Psi_{\rm act}\r = SG|\Psi_0\r\,.
\ee
We assume a general initial state
\be
\label{gc-init}
|\Psi_0\r=\sum_{\a,\b=0}^1c_{\a\b}|\a\b\r\,,
\ee
the minimisation in Eq.~(\ref{gc12}) runs over all possible
initial states $|\Psi_0\r$ and $\l\cdot\r$ denotes averaging over other
degrees of freedom of the system.
The imperfection in $G$ can be expressed
by an~{\it error operator} $E$ defined through
\be
\label{err-op1}
G = E G_{\rm perf} = E(S^{\dagger} P_{\bar{\vartheta}})\,,
\ee
where we used the relation $P_{\bar{\vartheta}}=SG_{\rm perf}$.
It follows from Eq.~(\ref{err-op1}) that
\be
\label{err-op2}
E
&=&
{\rm diag}\,\{
e^{i\delta\Theta_{00}},
e^{i\delta\Theta_{01}},
e^{i\delta\Theta_{10}},
e^{i\delta\Theta_{11}}
\}\,,\\[1mm]
\label{g-perf}
G_{\rm perf}
&=&
{\rm diag}\,\{
e^{i\bar{\Theta}_{00}},
e^{i\bar{\Theta}_{01}},
e^{i\bar{\Theta}_{10}},
e^{i\bar{\Theta}_{11}}
\}\,,
\ee
written in the~basis $\{|00\r, |01\r, |10\r, |11\r\}$
and we introduced the notation
\be
\label{gc16}
\delta\Theta_{\a\b}\equiv\Theta_{\a\b}-\bar{\Theta}_{\a\b}\,.
\ee
Bringing together Eq.~(\ref{gc-perf}), (\ref{gc-act}) and~(\ref{err-op1})
we obtain
\be
|\l\Psi_{\rm perf}|\Psi_{\rm act}\r|^2
&=&|\l\Psi_0|P_{\pi}^{\dag}SG|\Psi_0\r|^2\nonumber\\[1mm]
&=&|\l\Psi_0|P_{\pi}^{\dag}S(ES^{\dagger} P_{\bar{\vartheta}})|\Psi_0\r|^2
\nonumber\\[1mm]
&=& |\l\Psi_0|E|\Psi_0\r|^2   \,,   \label{err-op3}
\ee
where we used the commutation of diagonal operators and we assumed
$\bar{\vartheta}=\pi$ in order to accomplish the desired gate.
Hence the fidelity is
\be
\label{gc15}
{\cal F}
&=&
\left\l
\min_{\{c_{\a\b}\}}
\bigg|
\sum_{\a,\b=0}^1|c_{\a\b}|^2e^{i\delta\Theta_{\a\b}}
\bigg|^2
\right\r.
\ee
The~phases $\delta\Theta_{\a\b}$ are random variables and to obtain
a~single fidelity measure it is necessary to average
over characteristic probability distributions 
(associated with other degrees of freedom of the system).

\subsection{$\pi$-pulse method}
\label{pi}

The general concept in Eq.~(1)--(\ref{gc8})
of factoring a general phase rotation into
single-qubit and two-qubit terms can be applied to the error
operator $E$, giving
\be
\label{EZZ}
E=
Z_1(\delta \Theta_{00} - \delta \Theta_{10})\,
Z_2(\delta \Theta_{00} - \delta \Theta_{01})\,
P_{\delta\vartheta}\,,
\ee
where we refer to the notation in Eq.~(\ref{gc16}),
we denoted $\delta\vartheta=\vartheta-\bar{\vartheta}$
and we dropped a global phase
$(-\delta \Theta_{01}-\delta \Theta_{10})/2$.
We now consider a method which can be used to suppress
the single-qubit rotations in the~error operator~$E$,
leaving only the part which depends on
$\delta \vartheta$.
This method uses a pair of $\pi$ pulses on qubits~1 and~2 of the form
\be
\label{gc17}
R=R_1\otimes R_2\,,
\ee
where
\bse
\be
\label{gc18}
R_1 & = & |0\r_1\l 1|-|1\r_1\l 0|\,,\\
\label{gc19}
R_2 & = & |0\r_2\l 1|-|1\r_2\l 0|\,.
\ee
\ese
We replace the gate sequence $SG$ in Eq.~(\ref{gc10}) by $S^{\p}(RG)^2$,
to obtain
\be
\label{gc20}
\begin{array}{lcr}
|00\r & \stackrel{S^{\p}(RG)^2}{\ \ra\ } &
e^{i(\delta\Theta_{00}+\delta\Theta_{11})}\,|00\r\,,\\[1mm]
|01\r & \ra & e^{i(\delta\Theta_{01}+\delta\Theta_{10})}\,|01\r\,,\\[1mm]
|10\r & \ra & e^{i(\delta\Theta_{10}+\delta\Theta_{01})}\,|10\r\,,\\[1mm]
|11\r & \ra &
e^{i2\bar{\vartheta}}\,e^{i(\delta\Theta_{11}+\delta\Theta_{00})}\,|11\r\,,
\end{array}
\ee
where we used the notation (\ref{gc16}), $R$ is defined in
Eq.~(\ref{gc17}), $G$ is given by Eq.~(1) and
\be
\label{gc21}
S^{\p}=S^{\p}_1\otimes S^{\p}_2\,,
\ee
with
\be
\label{gc22}
S^{\p}_i
&=& Z_i (-\bar{\vartheta})
\ee
where $i=1,2$ and a global phase $\bar{\Theta}_{01}+\bar{\Theta}_{10}$ has
been dropped in Eq.~(\ref{gc20}).
Comparing Eq.~(\ref{gc10}) and (\ref{gc20}),
it~can be seen that
the joint logical state $|11\r$ is rotated during the new gate sequence
by $2 \bar{\vartheta}$, so now we require $2\bar{\vartheta} = \pi$.
This means that in the new sequence~(\ref{gc20})
each operation~$G$ will typically be applied for a time half as long as in
the~sequence~(\ref{gc10}).

\subsection{Fidelity of the phase gate using the~$\pi$-pulse method}
\label{fidwithpi}

When we use the gate sequence with the $\pi$ pulses in Eq.~(\ref{gc20}),
the actual state of the qubits is
\be
|\Psi_{\rm act}\r=S^{\p}(RG)^2|\Psi_0\r\,,
\ee
while the perfect state is given again by Eq.~(\ref{gc-perf}). Then,
the fidelity of the gate with the~perfect $\pi$~pulses turns out to be
\be
\label{gc24} 
{\cal F}'=
\left\l
\min_{\{c_{\a\b}\}} 
\bigg|
\sum_{\a,\b=0}^1|c_{\a\b}|^2e^{i(\delta\Theta_{\a\b}+\delta\Theta_{\ap\bp})}
\bigg|^2
\right\r,
\ee
where $\ap\equiv 1-\a$ and $\bp\equiv 1-\b$. 
We prove this in the~following
(note that Eq. (\ref{gc24}) and (\ref{fid-perf}) are
identical). Whenever $\delta\Theta_{\a\b}\ll 1$ and
$\delta\Theta_{\a\b}+\delta\Theta_{\ap\bp}\ll\delta\Theta_{\a\b}$,
we achieve a significant improvement in the~fidelity of the~phase
gate using the~$\pi$~pulses, that is $1-{\cal F}'\ll1-{\cal F}$.

The way the $\pi$-pulse method works can be brought out by
writing the gate sequence~(\ref{gc20}) using the error operator expressed
in Eq.~(\ref{EZZ}). That is
\be
\label{SRG}
S^{\p}(RG)^2
&=&
S^{\p}(R E G_{\rm perf})^2\nonumber\\[1mm]
&=&
S^{\p}\big[
R_1 R_2 Z_1(\phi) Z_2(\phi^{\p}) P_{\delta \vartheta} G_{\rm perf}
\big]^2,
\ee
where the phases $\phi$ and $\phi^{\p}$ are given in Eq.~(\ref{EZZ}), but
now we do not need to know what they are since we are about to prove that
they cancel.
Expressing the rotations using Pauli matrices, we have
\be
\label{rot1}
Z_j(\phi) &=& \cos(\phi/2) \openone_j + \sin(\phi/2) (\sigma_z)_j\,,\\[1mm]
\label{rot2}
R_j &=& i (\sigma_y)_j
\ee
for $j=1,2$.
Since $\sigma_y$ commutes with $\openone$ and anticommutes with $\sigma_z$,
we can write
\be
R_j\,Z_j(\phi)=Z_j(-\phi)\,R_j\,.  \label{RZ}
\ee
The operators $Z_j(\phi)$, $P_{\delta\vartheta}$ and $G_{\rm perf}$ all
commute, since they are diagonal, so the rotations $Z_j(\phi)$ cancel out
in Eq.~(\ref{SRG}) because they appear twice in the sequence with opposite
sign (after using Eq.~(\ref{RZ})). Then we get
\be
\label{pom1}
S^{\p}\left[
R_1 R_2 Z_1(\phi) Z_2(\phi^{\p}) P_{\delta\vartheta} G_{\rm perf}
\right]^2=
S^{\p}\left(R P_{\delta\vartheta} G_{\rm perf}\right)^2.
\nonumber\\
\ee
The fidelity of the phase gate using the~$\pi$-pulse method is
\be
\label{fprime}
{\cal F}' 
&=& 
\left\l
\min_{|\Psi_0\r} 
\bigg| 
\l\Psi_0|
\left[\big(G_{\rm perf}^{\dag} R^{\dag}\big)^2 \big(S^{\p}\big)^{\dag} \right]
\left[ S^{\p}\big(RG\big)^2 \right]
|\Psi_0\r 
\bigg|^2
\right\r .\nonumber\\
\ee
Using Eq.~(\ref{SRG}) and (\ref{pom1}) this becomes
\be
{\cal F}' &=& 
\left\l
\min_{|\Psi_0\r}
\left|
\l\Psi_0| \big(G_{\rm perf}^{\dag} R^{\dag}\big)^2
\big(R P_{\delta\vartheta} G_{\rm perf}\big)^2 |\Psi_0\r
\right|^2
\right\r\nonumber\\[2mm]
&=&
\left\l
\min_{|\Psi_0\r}
\left|
\l\Psi_0| G_{\rm perf}^{\dag}
\big( R P_{\delta\vartheta} \big)^2 G_{\rm perf} |\Psi_0\r
\right|^2
\right\r,
\ee
where we applied $R^{\dag}=R$. Let us introduce
\be
\label{err-opp}
E^{\p}\equiv
\big( R P_{\delta\vartheta} \big)^2
={\rm diag}\{e^{i\delta\vartheta}, 1, 1, e^{i\delta\vartheta}\}\,.
\ee
Hence, the error operator $E^{\p}$ is diagonal so it commutes
with $G_{\rm perf}$, giving
\be
\label{fid-perf}
{\cal F}'= 
\left\l
\min\left| \l\Psi_0 | E^{\p} | \Psi_0\r \right|^2
\right\r.
\ee
Clearly, when the uncertainties in the phases are such that the combination
$\vartheta$ has an uncertainty $\delta \vartheta$ small compared to the
uncertainties in the individual phases
$\delta \Theta_{\alpha\beta}$ in Eq.~(\ref{err-op2}),
then the $\pi$-pulse method offers a significant improvement
in the fidelity, that is $1-{\cal F}'\ll 1-{\cal F}$.

The fidelity of diagonal unitary matrices is discussed in
Appendix~\ref{E}. We will be interested in the case where
$\delta \Theta_{01} \simeq - \delta \Theta_{10} \gg
\delta \Theta_{00}, \delta \Theta_{11}.$
In this case the fidelity of the phase gate without
the~$\pi$~pulses is given by Eq.~(\ref{err-op3}) and~(\ref{fct}) as
\be
\label{Fid}
{\cal F}\simeq
\cos^2\big[(\delta\Theta_{01} - \delta\Theta_{10})/2\big]\,,
\ee
while for the phase gate using the~$\pi$-method Eq.~(\ref{fid-perf})
and~(\ref{f10}) give
\be
\label{Fp}
{\cal F}'\simeq\cos^2(\delta\vartheta/2)\,.
\ee
It should be recalled that each operation~$G$ is 
typically applied for half
the time in the gate sequence~(\ref{gc20}) compared to
the gate sequence~(\ref{gc10}). Therefore, the~quantities
$\delta\Theta_{\a\b}$ in Eq.~(\ref{Fid}) are not the same as
the quantities~$\delta \Theta_{\alpha \beta}$ in Eq.~(\ref{Fp}) as
the former are typically twice the size of the~latter.

The $\pi$-pulse method (``spin-echo'') to suppress an unwanted term
in a propagator (evolution operator) is one of
the standard tools of NMR spectroscopy,
which may be summarised as the observation that
\be
[\sigma_x Z(\phi)]^2 = \sigma_x Z(\phi) Z(-\phi) \sigma_x = \openone\,.
\ee
It relies on the assumption that
the rotations $Z$ in the~error operator $E$ in Eq.~(\ref{EZZ}) are
the~same in two successive
realisations of the gate operation $G$ in the sequence~(\ref{gc20}).
In the terminology of error
correction, it takes advantage of a known correlation
between two errors, thus giving an example of the fact that
correlation in noise can be advantageous.

\subsection{Imperfect $\pi$ pulses}
\label{imperfect}

The $\pi$-pulse method can be a very helpful tool and it~can strongly
suppress the infidelity of a computational process. However, it assumes
a cancellation of a large correlated error, which will only be exact when the
correlated error terms are identical and
the $\pi$~pulses are perfect. Such a cancellation will in practice be
sensitive to imprecisions in the~$\pi$~pulses and changes in
the~phases~$\Theta_{\a\b}$. We will next quantify this sensitivity.

The effect of a change in the phases ${\Theta}_{\a\b}$ between
the~two operations~$G$ in Eq.~(\ref{gc20}) can be absorbed
into the~imprecision of the~$\pi$~pulses.
The imprecision in the~$\pi$~pulses can be modelled as
a combination of error operators in the Pauli group. For instance,
the~imprecision in the~duration of a~$\pi$~pulse will give
an~``over-rotation" error of the form
\be
M &=& \left[
\begin{array}{rr}
\cos(\varepsilon/2) & -\sin(\varepsilon/2)\\
\sin(\varepsilon/2) & \cos(\varepsilon/2)
\end{array} \right] \nonumber\\[2mm]
&=&
\cos(\varepsilon/2)\,\openone-i\sin(\varepsilon/2)\,\sigma_y\,,
\label{gc25}
\ee
where $\varepsilon$ is the imprecision in the angle through which the~qubit
is rotated in its Bloch sphere, $\openone$ is the unity operator and $\sigma_y$
is the Pauli operator. More generally, the combination of the Pauli operators in
the dynamics will depend on the character of noise in the system and
may consist of
(i)~{\it unitary rotations}
(which may or may not be factorisable into a pair of
single-qubit rotations) or (ii)~a~{\it non-unitary relaxation}.
All these cases can be treated by using appropriate superpositions or
mixtures of the Pauli error operators. We analyse next examples of
noise of both types.

\subsubsection{Unitary rotations}

It is convenient to consider a general unitary error in the $\pi$
pulses as a combination of ``over-rotations'' in Eq.~(\ref{gc25}) and
phase errors corresponding to rotations about the~$z$~axis of
the Bloch sphere (see Eq.~(\ref{Zrot})).

First, we will consider pure phase errors. Let us suppose
the first $\pi$ pulse applied
to ion $j$ adds an additional rotation about the $z$ axis by
$\tilde{\varepsilon}_j$, and the second $\pi$ pulse a~rotation by
$p_j\tilde{\varepsilon}_j$, where $-1 \le p_j \le 1$. To analyse the~effects,
we replace the two occurrences of $R$ in Eq.~(\ref{SRG}) by
$R_1 R_2 Z_1(\tilde{\varepsilon}_1) Z_2(\tilde{\varepsilon}_2)$ and
$R_1 R_2 Z_1(p_1\tilde{\varepsilon}_1) Z_2(p_2\tilde{\varepsilon}_2)$, 
respectively. Using the commutation and anticommutation properties as
in Eq.~(\ref{rot1})--(\ref{fid-perf}),
the fidelity of the phase gate with the~imperfect $\pi$~pulses 
is found to be
\be
\label{Ftilde1}
\tilde{\cal F}
&=&
\left\l
\min\left| \l\Psi_0| Z_1\big[(1-p_1)\tilde{\varepsilon}_1\big]
Z_2\big[(1-p_2)\tilde{\varepsilon}_2\big]\, E^{\p} |\Psi_0\r\right|^2
\right\r
\nonumber\\[1mm]
&\simeq&
\left\l
1 - {\cal O}\left(
\big[\delta\vartheta + (1-p)\tilde{\varepsilon}\big]^2
\right)
\right\r.
\ee

Next, consider pure ``over-rotation'' errors. To simplify the algebra
we will assume both qubits have the same ``over-rotation'' angle, but
the two different $\pi$~pulses can have different ``over-rotation'' angles
$\varepsilon$ and $p \varepsilon$. Nothing essential is omitted by this
simplification. It is found that
\be
\label{Ftilde2}
\tilde{\cal F}\simeq
\big[1 - {\cal O}\left(\varepsilon^2\right)\big]{\cal F}'\,,
\ee
in which the coefficient $\varepsilon^2$ is of order $1 + p^2$
and ${\cal F}'$ is the~fidelity of
the phase gate with perfect $\pi$~pulses (see Appendix~\ref{F}).

\subsubsection{Non-unitary relaxation}

The operator $R$ in Eq.~(\ref{gc17}) represents perfect $\pi$~pulses.
Let us represent the~imperfect $\pi$~pulses with the following transformation
\be
\label{gc26}
{\cal R}(\rho)=\sum_{\mu=1}^4 M_{\mu}{\rho}M_{\mu}^{\dag}\,,
\ee
where $\rho$ is the density operator describing the state of
the qubits and $M_{\mu}$ are Lindblad operators of the form
\bse
\label{gc27}
\be
M_1&=&\sqrt{(1-\zeta_1)(1-\zeta_2)}\ R_1\otimes R_2\,,\\[1mm]
M_2&=&\sqrt{(1-\zeta_1)\zeta_2}\ R_1\otimes\openone_2\,,\\[1mm]
M_3&=&\sqrt{\zeta_1(1-\zeta_2)}\ \openone_1\otimes R_2\,,\\[1mm]
M_4&=&\sqrt{\zeta_1\zeta_2}\ \openone_1\otimes\openone_2\,,
\ee
\ese
which satisfy the condition
\be
\label{gc28}
\sum_{\mu=1}^4 M_{\mu}^{\dag} M_{\mu}=\openone\,.
\ee
The quantities
$\zeta_1$ and $\zeta_2$ are bit-flip error probabilities
for the qubit~1 and 2.
In other words, the operator $M_1$ represents the situation when both qubits
are successfully flipped, the operator $M_2$ ($M_3$) corresponds
to the failure to flip the qubit~2 (qubit~1) and finally the operator $M_4$
stands for the failure to flip both qubits.

When using the non-unitary relaxation (\ref{gc26}) to model errors due to
the~imperfect $\pi$~pulses we describe a~state of the qubits with
a~density operator. Then, the complete gate sequence reads
\be
\label{gc29}
\begin{array}{ccl}
\rho_0
&\stackrel{G}{\longrightarrow}&
G\rho_0G^{\dag}\\[4mm]
&\stackrel{S}{\longrightarrow}&
SG\rho_0G^{\dag}S^{\dag}\\[4mm]
&\stackrel{{\cal R}}{\longrightarrow}&
{\cal R}\left(SG\rho_0G^{\dag}S^{\dag}\right)\\[4mm]
&\stackrel{G}{\longrightarrow}&
G\left[{\cal R}\left(SG\rho_0G^{\dag}S^{\dag}\right)\right]G^{\dag}\\[4mm]
&\stackrel{\tilde{S}}{\longrightarrow}&
\tilde{S}G\left[{\cal R}\left(SG \rho_0 G^{\dag}S^{\dag}\right)\right]
G^{\dag}\tilde{S}^{\dag}\\[4mm]
&\stackrel{{\cal R}}{\longrightarrow}&
{\cal R}\left(\tilde{S}G
\left[{\cal R}\left(SG\rho_0G^{\dag}S^{\dag}\right)\right]
G^{\dag}\tilde{S}^{\dag}
\right),
\end{array}
\ee
where $\rho_0=|\Psi_0\r\l\Psi_0|$ denotes an initial state
of the~qubits. The single-qubit rotations $S$ are given by
Eq.~(\ref{gc4})--(\ref{gc5}) with the replacement~(\ref{gc9})
and we introduced additional single-qubit rotations in the form
\be
\label{gc30}
\tilde{S}=\tilde{S}_1\otimes\tilde{S}_2
\ee
with
\bse
\label{gc31}
\be
\tilde{S}_1&=&Z_1(\bar{\Theta}_{11}-\bar{\Theta}_{01})\,,\\
\tilde{S}_2&=&Z_2(\bar{\Theta}_{11}-\bar{\Theta}_{10})\,,
\ee
\ese
where we dropped a global phase
$\bar{\Theta}_{01}+\bar{\Theta}_{10}$ in Eq.~(\ref{gc29}).
The state of the qubits after the gate sequence~(\ref{gc29})
is derived in Appendix~\ref{A}.

The fidelity of the phase gate for an actual state described
by a density operator is defined as
\be
\label{gc33}
\tilde{\cal F}
=
\left\l
\min_{|\Psi_0\r}\,\l\Psi_{\rm perf}|\rho_{\rm act}|\Psi_{\rm perf}\r
\right\r,
\ee
where $|\Psi_{\rm perf}\r$ denotes the perfect state~(\ref{gc-perf}), while
$\rho_{\rm act}$ is the~actual state (\ref{a3}) after the gate sequence.
The~minimisation runs over all possible initial states~$|\Psi_0\r$.
Following Appendix~\ref{A} we obtain ($\zeta\ll 1$)
\be
\label{gc34}
\tilde{\cal F}=
(1-4\zeta){\cal F}'\,,
\ee
where $\tilde{\cal F}$ is the fidelity of the phase gate with
imperfect $\pi$~pulses and ${\cal F}'$ is the~fidelity
for perfect $\pi$~pulses calculated in Eq.~(\ref{gc24}).

Note that the results (\ref{Ftilde1}), (\ref{Ftilde2}) and
(\ref{gc34}) all have a~similar basic form, namely that
the~fidelity $\tilde{\cal F}$ is the~product of
the~fidelity ${\cal F}'$ for perfect
$\pi$~pulses and a quantity which is approximately
the fidelity of the~$\pi$~pulses themselves ($\zeta\sim\varepsilon^2$).
The essential point is that the system is not especially sensitive to
imperfection in the $\pi$ pulses, even though they are being used
to cancel a large correlated error.

\section{Ions in the microtraps}
\label{ions}

\begin{figure}[h]
\includegraphics[width=8cm]{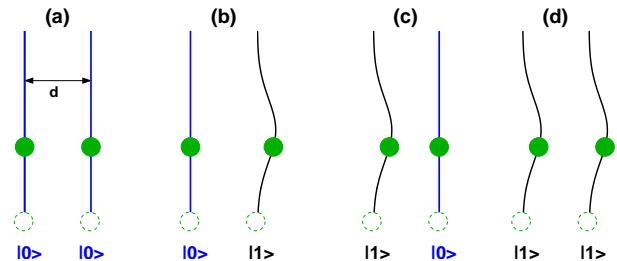}
\caption{The picture depicts trajectories of two ions corresponding to
internal states $|00\r, |01\r, |10\r, |11\r$, where a time-dependent force
applies on the ion only when it is in the internal state~$|1\r$.}
\label{config1}
\end{figure}

In this section we shall introduce the proposal of Cirac and Zoller for
the phase gate by pushing trapped ions in the~microtraps \cite{ions1,ions2}.
The basic principle is as follows. We consider two ions in two distinct
harmonic potentials (so~called {\it microtraps}) separated 
by a distance $d\simeq 1-500\mu{\rm m}$.
An internal-state-selective and time-dependent force is applied on both ions,
such that each ion experiences the~force
when it~is in an~internal (logical) state~$|1\r$ and no force
when it~is in an~internal (logical) state~$|0\r$ (FIG. \ref{config1}).

We assume that the pushing force is applied
on a time scale much longer than
an oscillation period of the ions in the microtraps (adiabatic
approximation). We also consider the displacement of the ions $\bar{x}$
due to the force to be small compared to the separation $d$
between the~microtraps ($\bar{x}\ll d$). Then the Coulomb repulsion energy
is given by
\be
E_{\rm coul}
&=&
\ell\left\{
\frac{1}{d},
\,\frac{1}{d+\bar{x}},
\,\frac{1}{d-\bar{x}},
\,\frac{1}{d}
\right\}\\[1mm]
&\approx&
\frac{\ell}{d} \left\{
1,
\,1 - \frac{\bar{x}}{d} + \left(\frac{\bar{x}}{d}\right)^2,
\,1 + \frac{\bar{x}}{d} + \left(\frac{\bar{x}}{d}\right)^2,
\,1
\right\}\nonumber
\ee
for the internal states $\{|00\r,|01\r,|10\r,|11\r\}$,
where $\ell=q^2/4\pi\epsilon_0$.
In this series expansion, the constant term produces
a~global phase. The term linear in $\bar{x}$ produces
single-qubit rotations, while the term quadratic in $\bar{x}$ produces
a two-qubit controlled-phase gate. If the force acts for time $\tau$,
then a phase gate with a phase $\vartheta$ expressed by Eq.~(\ref{gc8})
is obtained, where
\be
\vartheta
&\sim&
-\frac{\ell}{\hbar}\int_0^{\tau}
\left(
\frac{1}{d}-\frac{1}{d+\bar{x}}-\frac{1}{d-\bar{x}}+\frac{1}{d}
\right)dt\nonumber\\[1mm]
&\sim&
2\left(\frac{\bar{x}}{d}\right)^2
\frac{\ell\tau}{d \hbar}\,,
\ee
The single-qubit rotations which are produced at
the~same time are rotations through some angles $\phi$, 
where
\be
\phi\sim\frac{d}{\bar{x}}\vartheta\,.
\ee
More precise values of $\vartheta$ and $\phi$ will be obtained below.

The advantages of this gate are chiefly:
(i) the gate time can be short, and (ii) it is found that when a finite
temperature of the ions is taken into account, the dependence of
the phases on the motional state of the ions is small when $\bar{x} \ll d $.
However the condition $\bar{x} \ll d$
implies that the linear term in the Taylor expansion is large
compared to the quadratic term, and therefore when the~gate time
$\tau$ is large enough to produce the desired phase $\vartheta =
\pi$ or $\pi/2$, the Coulomb repulsion
produces a contribution to $\phi$ which is large compared to $\pi$
(typically $\phi\sim 500 \pi$).
This means that the single-qubit phases are
highly sensitive to small changes in the pushing force. Overall,
a~relative insensitivity to motional state is obtained at the~price
of sensitivity to fluctuations in the~pushing force. In Sec.~\ref{fluct}
we discuss further methods (in~addition to the $\pi$-pulse method)
to reduce the influence of these fluctuations.

\subsection{Hamiltonian}

The system of two ions trapped in two separate harmonic potentials
(microtraps) with a~time-dependent and
internal-state-selective force applied on the~ions
is described in the~semiclassical approach by the Hamiltonian
\be
\label{ion2}
H(t)=\sum_{\a,\b=0}^1
H_{\a\b}(t)\,|\a\r_1\l\a|\otimes|\b\r_2\l\b|\,,
\ee
with
\be
\label{ion3}
H_{\a\b}(t)
&=&
\frac{p_{\a}^2}{2m}+\frac{(p_{\b}^{\p})^2}{2m}\nonumber\\[1mm]
& &
+\frac{1}{2}m\omega^2(x_{\a}+d/2)^2+\frac{1}{2}m\omega^2(x_{\b}^{\p}-d/2)^2
\nonumber\\[1mm]
& &
+\big(s-x_{\a}-d/2\big)F_{\a}(t) + \big(s^{\p}-x^{\p}_{\b}+d/2\big)F_{\b}(t)
\nonumber\\[1mm]
& &
+\frac{\ell}{|x_{\b}^{\p}-x_{\a}|}\,,
\ee
where the two bare microtraps with a~trap frequency~$\omega$ 
are separated by
a~distance $d$, $m$ is the ion mass, $x_{\a}(t)$ and $x_{\b}^{\p}(t)$
are coordinates (trajectories) of ions~1 and~2
corresponding to their internal states
$|\a=0,1\r_1$ and $|\b=0,1\r_2$, $p_{\a}(t)$ and $p_{\b}^{\p}(t)$ are momenta
of the~ions. We denoted $\ell=q/4\pi\varepsilon_0$, where $q$ is 
the ion charge and $\varepsilon_0$ is the~permitivity of vacuum.
The~parameters~$s$ and $s^{\p}$ are associated with the potential which
the~ions experience when they are in their equilibrium positions~($x=\pm d/2$)
in the~microtraps. The quantity $F_{\a}(t)$ denotes
an~internal-state-selective and time-dependent force which displaces the ion
only when it is in its internal (logical) state~$|1\r$. 
We shall consider the~force in the form
\be
\label{ion4}
F_{\a}(t)=\a\hbar\omega f(t)/a\,,
\ee
where $\a=0,1$ refers to the internal ionic state $|0\r$ and $|1\r$,
$f(t)$ is a dimensionless time profile and we define
\be
\label{ion5}
a=\sqrt{\frac{\hbar}{m\omega}}\,.
\ee
The quantity $a$ is equal (up to the factor $1/\sqrt{2}$) to
the~ground state width of a quantum harmonic oscillator. In this
and subsequent sections it is assumed that the~force depends on time
but not on position. The effect of a dependence on position is then
discussed in Sec.~\ref{non-uni}.

\begin{figure}[h]
\includegraphics[width=8cm]{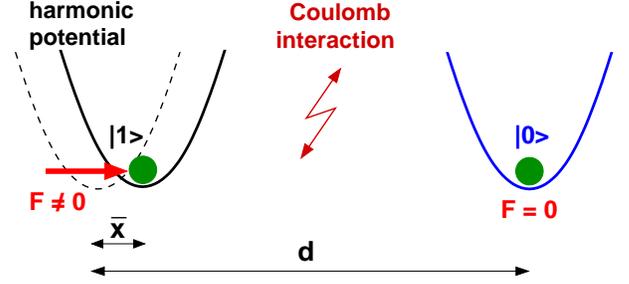}
\caption{Two ions (representing two qubits)
are trapped in two distinct harmonic potentials separated
by a~distance~$d$. A~state-selective force displaces
the trapping potential of the~ion only when it is in the logical state
$|1\r$.}
\label{trap}
\end{figure}

The coordinates $x_{\a}$ and $x_{\b}^{\p}$ describe positions of the~ions
with respect to the centre-of-mass of the system. It is convenient
to apply a~transformation
\bse
\label{ion6}
\be
x_{\a} &\rightarrow& x_{\a}-d/2\,,\\
x_{\b}^{\p} &\rightarrow& x_{\b}^{\p}+d/2\,,
\ee
\ese
where new coordinates refer rather to
the equilibrium positions of the ions in the microtraps.
When we use the~transformation~(\ref{ion6}), the~Hamiltonian~(\ref{ion3})
can be rewritten in the form
\be
\label{ion7}
H_{\a\b}(t)
&=&
\frac{p_{\a}^2}{2m}+\frac{(p_{\b}^{\p})^2}{2m}\nonumber\\[1mm]
& &
+\frac{1}{2}m\omega^2\left[
(x_{\a}-\bar{x}_{\a})^2 - \bar{x}_{\a}^2 + 2\bar{x}_{\a}s
\right]
\nonumber\\[1mm]
& &
+\frac{1}{2}m\omega^2\left[
(x_{\b}^{\p}-\bar{x}_{\b}^{\p})^2-(\bar{x}_{\b}^{\p})^2 +
2\bar{x}_{\b}^{\p}s^{\p}
\right]
\nonumber\\[1mm]
& &
+\frac{\ell}{|d+x_{\b}^{\p}-x_{\a}|}\,,
\ee
where we introduced
\bse
\label{ion8}
\be
\bar{x}_{\a}(t)&=&F_{\a}(t)/m\omega^2=\a a f(t)\,,\\
\bar{x}_{\b}^{\p}(t)&=&F_{\b}(t)/m\omega^2=\b a f(t)\,.
\ee
\ese
In Eq.~(\ref{ion7}) the action of the force
can be interpreted as a~displacement of the potential in which
each~ion is trapped and this displacement takes place only
when the ion is in its internal state~$|1\r$.
The~distance scale of the displacements is given by
the~quantity $a$ defined in Eq.~(\ref{ion5}).

\subsection{Dynamics}

The dynamics of the ion system is governed by the~evolution operator
\be
\label{ion9}
U=D\exp\left[
-\frac{i}{\hbar}\int_{t_0}^t H(t^{\p})\,dt^{\p}
\right]\,,
\ee
where the Hamiltonian is given by Eq.~(\ref{ion2}) and $D$~is the~Dyson
time-ordering operator. It follows from Eq.~(\ref{ion2}) and (\ref{ion9})
that
\be
\label{ion10}
U|\a\b\r=e^{i\Theta_{\a\b}}|\a\b\r\,,
\ee
where
\be
\label{ion11}
\Theta_{\a\b}=-\frac{1}{\hbar}\int_{t_0}^t H_{\a\b}(t^{\p})\,dt^{\p}\,.
\ee
Thus, the dynamics of the ion system described by
the~Hamiltonian~(\ref{ion3}) corresponds to a rotation of the~joint
internal state of the ions $|\a\b\r$ by {\it dynamical phases}
$\Theta_{\a\b}$. In Sec.~\ref{dp} we shall analyse the~phases~$\Theta_{\a\b}$
in order to apply them in the general discussion of Sec.~\ref{gc}.
However, first we have to understand the motion of the~ions in the microtraps.

\subsection{Motion of the ions}

Once we understand the motion of a single ion in a~microtrap with an external
pushing force, it is relatively simple to generalise its
behaviour to two ions because the~presence of a second ion
in a~separate microtrap causes via the Coulomb interaction
only small perturbations (under a~certain approximation)
to the motion of the first ion and vice versa.

The motion of a single trapped ion is discussed
in Appendix~\ref{C} and it appears that it consists of three different
motions. We see from
Eq.~(\ref{c7}) that the sloshing motion is a fast transitory motion
depending on the~turn-on rate of the force ($\dot{\bar{x}}\propto\dot{F}$).
In other words, when we turn on the~force sufficiently slowly compared to
oscillations of the~ion, the contribution of the sloshing part to
the overall motion will be negligible. Then, we refer to so called
{\it adiabatic approximation}. When we assume the displacement
of the potential and oscillations of the~ion to be of the~same order
($\bar{x}\sim\Delta$),
then the adiabatic approximation can be
estimated from the condition $|\delta x|\ll\min\{\bar{x},\Delta\}$ as follows
\be
\label{ion12}
\left|\int_{t_0}^t
\dot{f}(t^{\p})\,e^{i\omega t^{\p}}dt^{\p}
\right|\ll f(t)\,,
\ee
where $\dot{f}=df/dt$. For a Gaussian time profile of the force
\be
\label{ion13}
f(t)=\xi e^{-(t/\tau)^2},
\ee
the adiabatic approximation~(\ref{ion12}) requires
\be
\label{ion14}
e^{-(\omega\tau/2)^2}\ll 1\quad\Rightarrow\quad\omega\tau\gg 1\,.
\ee
This means that the time scale $\tau$ on which the force applies has
to be much longer than the oscillation period
$T_{\rm osc}=2\pi/\omega$ of 
the~ion. The duration of the force $\tau$ and the dimensionless
amplitude $\xi$ in Eq.~(\ref{ion13}) are
quantities which characterise the force. 
They have to be chosen such that the adiabatic condition~(\ref{ion14}) and
the~phase condition ($\bar{\vartheta}=\pi$ or $2\bar{\vartheta}=\pi$)
are satisfied.
In the adiabatic approximation the motion of the ion
which is in the~internal state $|\a\r$ simplifies to the form
\be
\label{ion15}
x_{\a}(t)\approx\bar{x}_{\a}(t)+\Delta(t)\,,
\ee
where the displacement $\bar{x}_{\a}(t)$ is defined in Eq.~(\ref{ion8}) and
the oscillations $\Delta (t)$ are given by Eq.~(\ref{c9}).

Now let us consider the case of two ions in two separate identical
traps. The Coulomb repulsion between the~ions affects their separation.
When two bare microtraps (i.e.~microtraps without the ions) are separated by
a distance $d$, then the equilibrium distance between two ions loaded
into the microtraps is bigger by the amount
\be
\label{ion19}
\Delta d=\frac{4d}{3}\sinh^2
\left\{\frac{1}{6}\ln\left[\eta+1+\sqrt{\eta(\eta+2)}\right]\right\}\,,
\ee
where $\eta=27\epsilon/4$ and
\be
\label{ion16}
\epsilon \equiv \frac{q^2}{\pi\varepsilon_0m\omega^2d^3}
\ee
gives the ratio between the Coulomb energy and the trapping
energy. For $\epsilon\ll 1$ the~expression in Eq.~(\ref{ion19})
simplifies to $\Delta d \approx \epsilon d/2$ but we will be also interested
in the~case when $\epsilon\lesssim 1$.

The presence of a second ion also causes
a~shift in the~oscillation frequency of the first ion and vice versa.
In a~3D treatment~\cite{ions2} and taking into account up to
the~quadratic term in the Coulomb energy,
the~oscillation frequency in the longitudinal direction
is corrected by the~factor
\be
\label{ion17}
{\omega_{\rm osc}}/{\omega_{\rm trap}}=\sqrt{1+\epsilon}\,,
\ee
with a factor $\sqrt{1-\epsilon/2}$ for the transversal oscillation
frequency. The higher order Coulomb terms cause the net potential
at each ion to be anharmonic and we will retain the third and fourth order
in the discussion to follow.
However, in order to keep the expressions uncluttered,
we will ignore the distinction between
$\omega_{\rm osc}$ and $\omega_{\rm trap}$.
In any case, a~fully accurate consideration
of this point requires a~fully quantum treatment~\cite{ions2}
of the motion rather than the semiclassical one adopted here.

\section{Dynamical phases}
\label{dp}

Now we are able to analyse in detail the dynamical phases $\Theta_{\a\b}$ in
Eq.~(\ref{ion11}) which have a key importance in
the quantum phase gate because they correspond
to the~phases in Eq.~(\ref{gc7}). We will assume
\begin{itemize}

\item[(i)] Adiabatic approximation -- given by Eq.~(\ref{ion14})
for the~Gaussian time profile of the force: ${\omega}\tau\gg 1$.

\item[(ii)] Displacements of the potentials and oscillations of the ions
are small compared to the trap separation: $\bar{x}, \Delta\ll d$
(weak force, low temperature).

\item[(iii)] Tight trapping potentials with respect
to the~Coulomb repulsion: $\epsilon\lesssim 1$.

\end{itemize}
Regarding the third assumption, the analysis simplifies
in the limit $\epsilon \ll 1$, and owing to this
some of the analytical results to be discussed will only be precise in that
limit. Namely, $\epsilon \ll 1$ allows us to assume the harmonicity of 
the potential what we used in Appendix~\ref{C}.
However, the~gate can still operate with high fidelity when
$\epsilon\sim 1$.

When we now substitute Eq.~(\ref{ion7})
into Eq.~(\ref{ion11}) we find that we deal with {\it single-particle}
and {\it two-particle} phases. Therefore, we write the phases
$\Theta_{\a\b}$ in the form
\be
\label{ion20}
\Theta_{\a\b}=\varphi_{\a}+\varphi_{\b}^{\p}+\phi_{\a}+\phi_{\b}^{\p}+
\phi_{\a\b}\,,
\ee
where $\a,\b=0,1$ and we introduce so called {\it kinetic phases}
\bse
\label{ion21}
\be
\varphi_{\a}&=&-\frac{1}{\hbar}\int_{t_0}^t
\frac{p_{\a}^2}{2m}\,dt^{\p}\,,\\
\varphi_{\b}^{\p}&=&-\frac{1}{\hbar}\int_{t_0}^t
\frac{(p_{\b}^{\p})^2}{2m}\,dt^{\p}\,,
\ee
\ese
then the single-particle phases due to the trapping potential
and the force
\bse
\label{ion22}
\be
\phi_{\a}&=&-\frac{1}{\hbar}\int_{t_0}^t
\frac{1}{2}m\omega^2\left[
(x_{\a}-\bar{x}_{\a})^2-\bar{x}_{\a}^2 + 2\bar{x}_{\a}s
\right]dt^{\p},\nonumber\\
\\
\phi_{\b}^{\p}&=&-\frac{1}{\hbar}\int_{t_0}^t
\frac{1}{2}m\omega^2\left[
(x_{\b}^{\p}-\bar{x}_{\b}^{\p})^2-(\bar{x}_{\b}^{\p})^2 + 2\bar{x}_{\b}^{\p}s'
\right]dt^{\p},\nonumber\\
\ee
\ese
and finally the two-particle phases originating from
the~Coulomb interaction between the ions
\be
\label{ion23}
\phi_{\a\b}=-\frac{1}{\hbar}\int_{t_0}^t
\frac{\ell}{|d+x_{\b}^{\p} - x_{\a}|}\,dt^{\p}\,.
\ee
We will assume that $t_0<0$, $t>0$ and $|t|,|t_0|\gg\tau$, where
$\tau$~is the gate time.

\subsection{Kinetic phases}

The kinetic phases in Eq.~(\ref{ion21}) are associated with the~kinetic
energy of the ions. We can calculate them using the
expression for the motion of the ions in Eq.~(\ref{ion15}). Taking into
account that $p=m\dot{x}$ and assuming that $|t_0|,|t|\to\infty$
(it means that we turn the~force gradually on from the zero value
and turn it slowly off back to the~zero) we can write
\be
\label{ion24}
\varphi_{\a}
&=&-\frac{m}{2\hbar}\int\limits_{-\infty}^{+\infty}\left(
\dot{\bar{x}}_{\a}^2+2\dot{\bar{x}}_{\a}\dot{\Delta}+\dot{\Delta}^2
\right)dt\nonumber\\[1mm]
&=&\varphi_{\a}^{I}+\varphi_{\a}^{I\!I}+\varphi^{I\!I\!I}\,.
\ee
For the Gaussian time profile~(\ref{ion13}) we get
\be
\label{ion25}
\varphi_{\a}^I=
-\frac{\a^2\xi^2}{\omega\tau}\sqrt{\frac{\pi}{8}}
\ee
and
\be
\label{ion26}
\varphi_{\a}^{I\!I}=-\a\omega\tau\xi e^{-(\omega\tau/\sqrt{2})^2}
\sqrt{\frac{2\pi\E}{\hbar\omega}}\cos\psi\,,
\ee
where $\a=0,1$ correspond to the internal states $|0\r$ and $|1\r$, $\xi$
and $\tau$ are defined in Eq.~(\ref{ion13}) and $\E$ with $\psi$ are
introduced in Eq.~(\ref{c9}).
The~contribution $\varphi^{I\!I\!I}$ in Eq.~(\ref{ion24})
is a~global phase which does not depend on the internal state $|\a\r$ and
can be omitted.

The displacement $\bar{x}$ scales with the~quantity~$a$ introduced
in Eq.~(\ref{ion5}), and by assumption~(ii) it is much smaller
than the ion separation $d$.
Therefore, it is reasonable to assume that
the~parameter~$\xi$ in Eq.~(\ref{ion8}) is of the~order
of one or smaller ($\xi\lesssim 1$).\label{xi} Then, it follows from
Eq.~(\ref{ion25}) and (\ref{ion26}) that $\varphi_{\a}^I<\pi$,
or $\varphi_{\a}^I\ll\pi$ when the~adiabatic condition is
very well obeyed.

The contribution $\varphi_{\a}^{I\!I}$ vanishes very rapidly
with the~adiabatic approximation~(\ref{ion14}).
Outside the adiabatic regime the phase $\varphi_{\a}^{I\!I}$ 
would be responsible for
a sensitivity of the gate to the motional state because
$\varphi_{\a}^{I\!I}$ scales with $\sqrt{\E}$, where $\E$ is the oscillation
energy of the~ion. In the full quantum approach we would find that
the~phase contribution $\varphi_{\a}^{I\!I}$ corresponds to entanglement
between motional and internal degrees of freedom of the~ions which
spoils the performance of the gate.

\subsection{Single-particle potential energy phases}

The phases in Eq.~(\ref{ion22}) are associated with the trapping potential
energy and energy introduced to the system via the external force $F$.
When we substitute Eq.~(\ref{ion15}) into Eq.~(\ref{ion22}) we
obtain
\be
\label{ion27}
\phi_{\a}&=&
\sqrt{\frac{\pi}{8}}\,\a\omega\tau\xi^2
- \sqrt{\pi}\,\a\omega\tau\xi\frac{s}{a}\,,
\ee
where $\a=0,1$ and we omitted the first term in Eq.~(\ref{ion22})
because it contributes only a~global phase. With
assumptions~(i) and~(ii),
where $\bar{x}\sim a$, we find $\phi_{\a}\gg\pi$.
We obtain the~expression for $\phi_{\b}^{\p}$ when we replace
$\a$ with $\b$ and $s$ with $s'$ in Eq.~(\ref{ion27}).

\subsection{Two-particle interaction phases}

Finally, we can analyse the two-particle interaction phases in
Eq.~(\ref{ion23}). They have a key importance in the~scheme of the phase
gate because they actually produce the phase $\vartheta$ in Eq.~(\ref{gc7}).
The phases $\phi_{\a\b}$ are highly nonlinear but we can expand them
in a Taylor series using assumption~(ii). Then we can rewrite
Eq.~(\ref{ion23}) in the~form
\be
\label{ion28}
\phi_{\a\b}=
-\frac{\ell}{\hbar d}\int_{t_0}^{t}\sum_{n=0}^{\infty}
\left(\frac{x_{\a}-x_{\b}^{\p}}{d}\right)^n dt^{\p}\,,
\ee
where $\phi_{\a\b}$ are calculated with precision up to $n=4$ in
Appendix~\ref{D}. The term of order~$n=0$
contributes only with a global phase and can be omitted. The linear term
($n=1$) in Eq.~(\ref{d5}) has a~character of a~single-qubit phase and
it is the~quadratic term ($n=2$) in Eq.~(\ref{d7}) which finally produces
a~two-particle interaction between the~ions via the Coulomb repulsion.
We consider also higher order terms with $n=3$ and $n=4$
in Eq.~(\ref{d8}) and~(\ref{d9})
because they contribute with thermal motion contributions.
In principle it would be sufficient to take into account only the cubic term
($n=3$) but with the~help of the~$\pi$-pulse method we are able to
undo contributions of
the~odd terms. Therefore, we have to consider also the~biquadratic term
($n=4$) as a lowest order correction to the quadratic term when we apply 
the~$\pi$-pulse method.

Using the results from Appendix~\ref{D} we can calculate
the~phase $\vartheta$ defined by Eq.~(\ref{gc8}). That is
\be
\label{ion29}
\vartheta
&=&\Theta_{11}-\Theta_{10}-\Theta_{01}+\Theta_{00}\\[1mm]
&=&\phi_{11}-\phi_{10}-\phi_{01}+\phi_{00}\nonumber\\[1mm]
&=&\theta\left\{
1+\left(\frac{a}{d}\right)^2\right.\times\nonumber\\[1mm]
& &
\times\left.\left[
\frac{\xi^2}{\sqrt{2}}+6\left(\frac{E_1}{\hbar\omega}+\frac{E_2}{\hbar\omega}
-2\frac{\sqrt{E_1E_2}}{\hbar\omega}\cos(\Delta\psi)\right)\right]
\right\}\,,\nonumber
\ee
where all single-particle phases
$\varphi_{\a}$, $\varphi_{\b}^{\p}$, $\phi_{\a}$, $\phi_{\b}^{\p}$
cancelled out and only the two-particle phases $\phi_{\a\b}$ contributed,
$\Delta\psi=\psi_1-\psi_2$
and we introduced the quantity
\be
\label{ion30}
\theta\equiv\sqrt{\frac{\pi}{8}}\,\epsilon\omega\tau\xi^2\,.
\ee
It follows from assumption~(i) that $\omega\tau\gg 1$, from
assumption~(iii) that $\epsilon\lesssim 1$ and we aim for the phase gate with
$\vartheta=\pi$ or $2\vartheta=\pi$. Therefore, we confirm
that $\xi \lesssim 1$ as we assumed previously in the~discussion
after Eq.~(\ref{ion26}).

In order to calculate the fidelity of the phase gate with ions in
the~microtraps we need to express the~quantity~$\delta\Theta_{\a\b}$
in Eq.~(\ref{gc16}). However, first we need to make a~good estimate for
the~phases~$\bar{\Theta}_{\a\b}$ in Eq.~(\ref{gc9}).

After laser cooling we can describe the oscillation motion of the ions
with the Boltzmann thermal distribution with the probability
\be
\label{ion31}
p(\E)=\frac{1}{Z}\,e^{-\E/k_BT}\,,
\ee
where $k_B$ is the Boltzmann constant, $T$ is the temperature and $\E$ is
the~oscillation energy of the ion.
Further, it~seems to be realistic to describe
the~oscillation phases $\psi_1$ and $\psi_2$ in Eq.~(\ref{d4}) with
a~uniform distribution. Then, a~good estimate for
the phases $\bar{\Theta}_{\a\b}$ is the average value
\be
\label{ion32}
\bar{\Theta}_{\a\b}\equiv
\int\limits_0^{2\pi}\frac{d\psi_1d\psi_2}{(2\pi)^2}
\int\limits_0^{+\infty}\frac{d\E_1d\E_2}{(k_BT)^2}
\ \Theta_{\a\b}\ e^{-(\E_1+\E_2)/k_BT}\,.\nonumber\\
\ee
Taking the case that fluctuations in the pushing force are
negligible, the~parameter~$\xi$ has a fixed value during
the~gate operation. Then, we can write
\be
\label{ion33}
\delta\Theta_{\a\b} =\Theta_{\a\b}-\bar{\Theta}_{\a\b}
                    =\phi_{\a\b}-\bar{\phi}_{\a\b}\,,
\ee
and the gate phase in Eq.~(\ref{gc11}) is
\be
\label{ion34}
\bar{\vartheta}
=\theta\left[1+
\left(
\frac{a}{d}\right)^2\left(\frac{\xi^2}{\sqrt{2}}+\frac{12k_BT}{\hbar\omega}
\right)
\right]\,,
\ee
where we used Eq.~(\ref{d11}).
We require $\bar{\vartheta}=\pi$ or $2\bar{\vartheta}=\pi$
to realise the phase gate and it follows from assumption~(ii)
that $a\ll d$ (because $\xi\lesssim 1$).
Therefore, the~phase condition reduces to $\theta\approx\pi$ or
$2\theta\approx\pi$. It follows from Eq.~(\ref{ion30}) that the phase
condition determines directly the~{\it gate time} $\tau$ (for given
values of the other parameters).

Finally, we can calculate the fidelity of the phase gate without
the~$\pi$~pulses given by the sequence~(\ref{gc7}).
When we substitute Eq.~(\ref{d10}) and (\ref{d11})
into Eq.~(\ref{gc15}) we get
\be
\label{ion35}
{\cal F}=1-\left(\frac{6\theta k_BT}{\hbar\omega}\right)^2
\left[
\frac{1}{\xi^2}\left(\frac{a}{d}\right)^2-2\left(\frac{a}{d}\right)^4
\right]\,,
\ee
where the minimisation is carried out in the original paper~\cite{ions2} and
this result corresponds also to Eq.~(\ref{Fid}).
The~fidelity of the phase gate with perfect $\pi$~pulses defined by
Eq.~(\ref{gc24}) and corresponding to the sequence~(\ref{gc20})
is
\be
\label{ion36}
{\cal F}'=
1-\left(\frac{6\theta k_BT}{\hbar\omega}\right)^2
\left(\frac{a}{d}\right)^4\,,
\ee
corresponding to Eq.~(\ref{Fp}).
Comparing Eq.~(\ref{ion35}) and (\ref{ion36}) it is seen that
the~$\pi$-pulse method improves the fidelity by two orders in $a/d$, and we will
see later that $a/d\sim 10^{-3}$.

\section{Pushing force}
\label{force}

The state-dependent force was
introduced in Eq.~(\ref{ion3}) and Eq.~(\ref{ion4}), but
we have not yet specified how it~can be realised in practice.
In this section we discuss
the~dipole force (originating from a laser beam) as a~suitable
candidate force. We will consider the dipole force in
a~travelling-wave and a standing-wave configuration.
The dipole force is always associated with some unwanted photon
scattering which reduces the fidelity of the gate. Quantifying
this is the main aim of this section.

\subsection{Dipole force}
\label{dipole}

\begin{figure}[h]
\includegraphics[width=8cm]{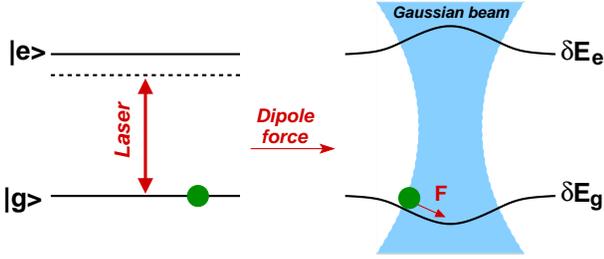}
\caption{Non-resonant laser beam causes AC Stark light shifts with
a magnitude dependent on the intensity profile of the~laser beam.
The light shifts produce a dipole force which pushes the atomic
particle towards the minimum of the laser-light potential.}
\label{dipole2}
\end{figure}

The {\it dipole force} is produced in an intensity gradient of
the light illuminating the atom, typically
a~laser beam which is far detuned from some atomic transition.
The~relevant atomic levels are light-shifted (AC~Stark shift) creating
an~additional potential for the particle. When the laser is tuned below
the~atomic frequency (red detuning), the particle is driven
towards the laser intensity maximum, i.e. to the minimum of the created
potential (FIG.~\ref{dipole2}).
For a~blue detuning the particle is pushed away from the~intensity maximum.

\begin{figure}[h]
\includegraphics[width=8cm]{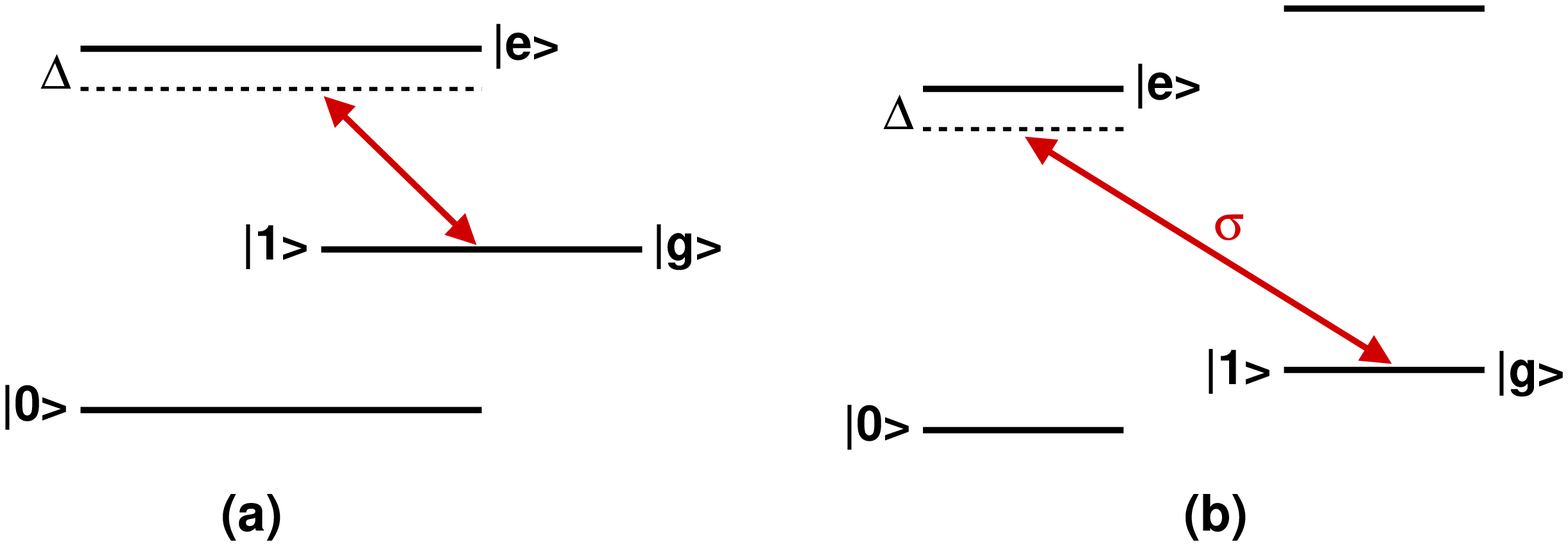}
\caption{(a) Logical states $|0\r$ and $|1\r$ of the ion qubit are
represented by a ground and excited (metastable) atomic (ionic)
levels. A~non-resonant laser does not couple the state $|0\r$. 
(b)~Logical states are represented by two sublevels of a ground state
manifold. Coupling of the $|0\r$ state is prevented by
polarization selection rules, for example using a
$\sigma$-polarized laser.} \label{detun}
\end{figure}

Let us assume a~two-level atom (ion) with a~lower level~$|g\r$ and
an~upper level~$|e\r$ separated by an~atomic frequency~$\omega_A$.
The~$|g\r$ state corresponds to the logical state $|1\r$ of the ion while
the $|e\r$ state is an excited auxiliary state. The logical
state $|0\r$ is not to be coupled to the light (FIG.~\ref{detun}).
When we apply a~laser beam of frequency $\omega_L$
the atom-laser Hamiltonian in the dipole approximation is
\be
\label{force2}
H=\frac{\hbar\omega_0}{2}\sigma_z+
\frac{\hbar\Omega}{2}(\sigma_{+}+\sigma_{-})
\left[e^{i(\omega_Lt+\phi_L)}+e^{-i(\omega_Lt+\phi_L)}\right]\!,
\nonumber\\
\ee
where $\sigma_z=|e\r\l e|-|g\r\l g|$, $\sigma_{+}=|e\r\l g|$,
$\sigma_{-}=|g\r\l e|$,
$\Omega(\bs{r},t)$ is the Rabi frequency on the transition $|e\r \leftrightarrow |g\r$
and we assumed the~laser can be treated as a~classical light field.
In the limit of large detuning $\Delta=\omega_L-\omega_A\gg|\Omega|$,
the standard treatment of this ``two-level atom'' leads
to the dipole force (on the atom in the~$|g\r$~state)
\be
\label{force15}
\bs{F}_{\rm dip}= 
-\frac{\hbar}{4\Delta}\bs{\nabla}
\left|\Omega(\bs{r})\right|^2.
\ee

In the {\it travelling-wave configuration}
we place the~ion off the axis of the~laser beam
and use its non-uniform intensity profile
\be
\label{force16}
\Omega_{\rm trav}=
\Omega_0\,e^{-(t/\tau\sqrt{2})^2}e^{-[(x-x_0)/w]^2}e^{ikz}\,,
\ee
where $x-x_0$ refers to the position of the ion in the~profile (cross-section)
of the laser beam.
We assumed the~exponential time profile of the laser pulse with
a duration $\tau$ and the Gaussian intensity profile with the size of
the~beam waist $w$ and the radial distance $x$.
We also introduced the constant
\be
\label{force17}
|\Omega_0|^2=\frac{6\pi\Gamma I}{\hbar c k^3}
=\frac{12\Gamma P}{\hbar c k^3 w^2}\,,
\ee
where $\Gamma$ is the line width of the $|g\r\rightarrow |e\r$ transition,
$I$ and $P$ are the intensity and power of the laser, $c$ is the~speed of
light and $k=2\pi/\lambda$ is the wave number with the~laser
wavelength $\lambda$.

\begin{figure}[h]
\includegraphics[width=8.5cm]{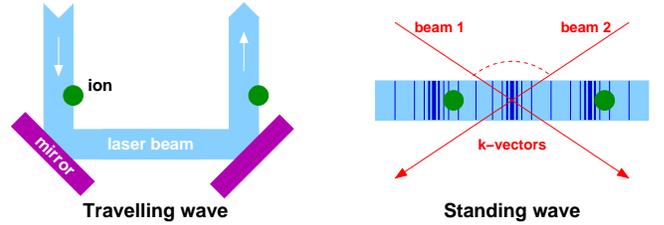}
\caption{Scheme of the travelling-wave and the standing-wave configuration.
The travelling wave is created by a single laser beam deflected by mirrors.
The standing wave is created by two laser beams with different $k$-vectors.
By adjusting the~angle between the $k$-vectors the period of the standing
wave (node-antinode separation) can be changed.}
\label{config2}
\end{figure}

In the {\it standing-wave configuration} we rather use the~gradient of
the standing field of the laser, giving
\be
\label{force18}
\Omega_{\rm stan}=
2i\Omega_0\,e^{-(t/\tau\sqrt{2})^2}e^{-(x/w)^2}\sin[k(z-z_0)]\,,
\ee
where $z-z_0$ corresponds to the position of the ion in the~standing-wave field.
When we now substitute Eq.~(\ref{force17}) and (\ref{force18}) into
Eq.~(\ref{force15}) we calculate the pushing dipole force on the~ions.
Comparing the expression of the dipole force~(\ref{force15})
with the general form
of the pushing force in Eq.~(\ref{ion4}) we can express
the dimensionless amplitude $\xi$, giving
\bse
\label{force19}
\be
\xi_{\rm trav}&=&
\frac{a(x-x_0)}{\omega\Delta w^2}\,|\Omega_0|^2\,e^{-2[(x-x_0)/w]^2}\!,\\[1mm]
\xi_{\rm stan}&=&
-\frac{ak}{\omega\Delta}\,|\Omega_0|^2\,\sin[2k(z-z_0)]\,e^{-2(x/w)^2}\!.
\ee
\ese
It follows from Eq.~(\ref{force19}) that the maximum force
is for $x_0=w/2$ in the travelling-wave and for
$kz_0=\pi/4$ in the~standing-wave configuration.

\subsection{Photon scattering}
\label{scat}

We use a far detuned laser to create a dipole pushing force on the ions.
Even though the laser is far detuned from the $|g\r\leftrightarrow|e\r$
transition there is still some photon scattering due to a non-zero
line width $\Gamma$ of the transition. A single scattered photon is
sufficient to destroy the~fidelity of the gate, because scattering only
takes place in the qubit logical state $|1\r$,
not in the~logical state $|0\r$. Therefore,
a~scattering event constitutes a measurement of the qubit by the
environment. We have to choose parameters so that the number of photons
scattered during the gate is small compared to one ($N\ll 1$).

The {\it scattering rate} in the two-level atom approximation and 
in the~weak-coupling regime~($|\Omega|,\Gamma\ll|\Delta|$) is~\cite{book}
\be
\label{force21}
R=\frac{\Gamma}{2}\frac{|\Omega|^2/2}{\Delta^2+(\Gamma/2)^2+|\Omega|^2/2}
\approx\frac{\Gamma}{4}\frac{|\Omega|^2}{\Delta^2}\,.
\ee
The {\it number of scattered photons} during the gate operation, i.e. during
the period when the laser is on and drives the $|g\r\leftrightarrow|e\r$
transition, is given by
\be
\label{force22}
N=\int_{t_0}^t R(t')\,dt'\,,
\ee
where we assume that the laser is gradually switched on and off
as stated in Eq.~(\ref{force16}) and (\ref{force18}), and
the~comment below Eq.~(\ref{ion23}) applies.
When we now substitute for the~coupling constant~$\Omega$ the number of
scattered photons can be expressed in the~travelling-wave configuration as
\be
\label{force24}
N_{\rm trav}\approx
C_{\rm trav}\,\frac{w^6}{x_0^2}\frac{d^{\,3}\omega^4}{P}\,
e^{2(x_0/w)^2}
\ee
and in the standing-wave configuration as
\be
\label{force25}
N_{\rm stan}\approx
C_{\rm stan}\,w^2\,\frac{d^{\,3}\omega^4}{P}
\frac{1}{\cos^2(kz_0)}\,,
\ee
where the constants are
\bse
\label{force26}
\be
C_{\rm trav}
&=&
\left(\frac{\pi^5\sqrt{2}}{3}\frac{\varepsilon_0 c}{q^2}\right)
\frac{m^2}{\lambda^3}\,,\\[1mm]
C_{\rm stan}
&=&
\left(\frac{\pi^3\sqrt{2}}{12}\frac{\varepsilon_0 c}{q^2}\right)
\frac{m^2}{\lambda}\,,
\ee
\ese
and we assumed that the distance moved by the ion was small compared to 
the~waist size~$w$ (period $\pi/k$) in the~travelling-wave (standing-wave)
configuration.

To calculate the number of scattered photons in Eq.~(\ref{force24})
and~(\ref{force25}) we also needed to express the gate time $\tau$
from Eq.~(\ref{ion30}), where the~phase condition for the~gate with
the~$\pi$~pulses has to be satisfied. It follows from Eq.~(\ref{ion34}) that
$2\theta\approx\pi$, giving
\be
\label{force27}
\tau=\frac{\sqrt{2\pi}}{\epsilon\omega\xi^2}\,,
\ee
and we use the expression in Eq.~(\ref{force19})
for the parameter~$\xi$. We also took into account that we apply two
successive gate pulses.

When we compare photon scattering for the travelling-wave and the standing-wave
configuration we get (assuming $x_0\sim w$)
\be
\label{force28}
\frac{N_{\rm trav}}{N_{\rm stan}}\sim\left(\frac{w}{\lambda}\right)^2\,,
\ee
where $w$ is the size of the laser beam waist and $\lambda$ is
the~wavelength of the laser light. In practice, we shall have
$w\sim 10^{-6}\,{\rm m}$ and $\lambda\sim 10^{-7}\,{\rm m}$, that is
$N_{\rm trav}/N_{\rm stan}\sim 100$. This means that in the travelling-wave
configuration we have approximately 100-times more photon scattering. 
This is because for a given laser intensity,
the smaller intensity gradient of the travelling wave 
(compared to the~gradient of the~standing wave)
leads to a~smaller force, hence a~longer gate time and more scattered photons.
Alternatively, if the force is increased for the travelling wave case
by reducing the detuning, the~scattering rate goes up as $1/\Delta^2$ 
and hence again there are more scattered photons.

The travelling-wave configuration nevertheless has an~advantage.
The ions oscillate in the microtraps due to non-zero temperature and
experience different instant magnitudes of the force in different points
within the laser beam. In the travelling-wave configuration
the excursion of the ion thermal motion $\Delta_{\rm osc}$ is limited 
approximately by $\Delta_{\rm osc}\ll w$, 
while in the standing-wave configuration it~has the stronger
constraint $\Delta_{\rm osc}\ll\lambda$ (Lamb-Dicke regime). Therefore, the~latter
case requires more laser cooling and it is generally more experimentally
demanding.

\subsection{Contribution of scattering to the infidelity}

We have already discussed in Sec.~\ref{scat} that photon scattering appears
only when the ion is in the logical state $|1\r$ and therefore constitutes
a~measurement of the~qubit. A measurement will cause a state
reduction such that fidelity falls to $1/2$ in the worst case.
However, depending on the atomic structure, photon scattering
might also be associated with optical pumping of the state from the~state 
$\ket{1}$ to $\ket{0}$, in which case a single scattered photon can reduce 
the~fidelity to zero in the worst case. Making the latter (i.e. more cautious)
assumption, we therefore estimate the
influence of photon scattering on the fidelity as
\be
\label{infid3}
{\cal F}_{\rm scat}=e^{-N}\approx 1-N\,,
\ee
where $N$ is the overall number of scattered photons during the phase gate
and it only makes sense to consider $N\ll 1$. The number $N$ is given by
Eq.~(\ref{force24}) and~(\ref{force25}) depending on the experimental
realisation.

\section{Influence of Fluctuations}
\label{fluct}

\subsection{Fluctuations in single-qubit phases}
\label{fluct2}

In the original work \cite{ions2} the single-particle phases were
regarded as not contributing to loss of fidelity, by the~argument
that as long as their value is known, the correct single-qubit
rotation operator can be applied to a~very good precision. However,
as we commented in Sec.~\ref{dp}, there are present some very large terms
in the single-qubit phases, so
that even a small relative change in the~value of a parameter may
introduce an unknown phase
change of the order of $\pi$ and completely destroy the fidelity.
For example, when the force is provided by a laser beam as
discussed in Sec.~\ref{force}, intensity fluctuations of the laser
cause fluctuations in the pushing force~$F$ and therefore
in its amplitude~$\xi$. The single-qubit
phases are highly sensitive to such changes when $a\ll d$
(or more precisely $a\xi\ll d$).

Bringing together Eq.~(\ref{ion20}), (\ref{ion25}), (\ref{ion27}) and
(\ref{d11}) we obtain the phase required for the single-qubit
operation~$S_1$ in Eq. (\ref{gc5}), giving
\be
\label{Phi1}
\Phi_1
&\equiv&
\bar{\Theta}_{10} - \bar{\Theta}_{00}\nonumber\\[1mm]
&=&
\theta
\left[
-\frac{1}{\epsilon(\omega\tau)^2}
+\frac{1}{\epsilon}
-\frac{s}{a}\frac{\sqrt{8}}{\xi\epsilon}
-\frac{d}{a}\frac{1}{\xi\sqrt{2}}
-\frac{1}{2}
+{\cal O}(a/d)
\right],\nonumber\\
\ee
where the first term corresponds to
the~kinetic phase $\varphi_{\a}^I$ in Eq.~(\ref{ion25}) and we dropped
$\varphi_{\a}^{I\!I}$ since it~is strongly suppressed in the adiabatic
limit. The~second and the~third term come from the potential energy phase
in Eq.~(\ref{ion27}) and they correspond to the {\it light shift phases}
because the~pushing force is represented by the dipole force.
The~next two terms correspond to the linear and quadratic
contribution from the Taylor expansion of the~Coulomb energy and
${\cal O}(a/d)$ denotes higher order terms in this expansion.

The expression for
$\Phi_2 \equiv \bar{\Theta}_{01} -\bar{\Theta}_{00}$ is
the same as $\Phi_1$ in Eq.~(\ref{Phi1}) except $s$ is replaced by $s'$ and
the odd-order Coulomb terms in~$d/a$ and $a/d$ change sign.
Further, we will discuss the case when $\Phi_j$ can be different
in two successive gate operations $G$ in the sequence~(\ref{gc20})
due to the influence of fluctuations of experimental parameters.

The $\pi$-pulse method goes some way towards alleviating
the~effects of these fluctuations, since then the overall sequence is
only sensitive to a change in the single-qubit phases between
the two successive applications of the gate operation~$G$.
Let these two successive
pushing gates be $G_1$ and $G_2$, so the gate sequence for
the phase gate with the~$\pi$~pulses is $S'(R G_2) (R G_1)$.
Using Eq.~(\ref{err-op1}) we define error operators
\be
\label{fl1}
E_j\equiv G_jG^{\dag}_{\rm perf}\,,
\ee
where $j=1,2$. Then, the gate sequence is
\be
\label{fl2}
S'(RG_2)(RG_1)
\ \rightarrow\
S'\big[R(E_2G_{\rm perf})\big]\big[R(E_1G_{\rm perf})\big]\,.\nonumber\\
\ee
Without loss of generality,
the error operator $E_1$ may be taken equal to $E$ in Eq. (\ref{EZZ})
and we write $E_2$ in the form
\be
\label{fl3}
E_2=
Z_1(\Delta\Phi_1)\,
Z_2(\Delta\Phi_2 )\,
P_{\Delta\bar{\vartheta}}\,E
\ee
where $\Delta\Phi_j$ are the changes (due to fluctuations) 
in $\Phi_j$ between the two successive applications of the pushing gate, 
that is
\be
\label{fl4}
\Delta\Phi_j=\left(\Phi_j\right)_{G_2}-\left(\Phi_j\right)_{G_1},
\ee
while $\Delta\bar{\vartheta}$ is the change in $\bar{\vartheta}$, that is
\be
\label{fl5}
\Delta\bar{\vartheta}=
\bar{\vartheta}_{G_2}-\bar{\vartheta}_{G_1}.
\ee
Notice that for $\Delta\Phi_j=0$ and $\Delta\bar{\vartheta}=0$
(no fluctuations), the sequence~(\ref{fl2}) corresponds
to the sequence~(\ref{SRG})
and both error operators $E_1$ and $E_2$ are equal to $E$ in
Eq.~(\ref{err-op1}).

The~imprecision represented by $\Delta\bar{\vartheta}$
is similar in its
effects to those of $\delta\vartheta$ because it reduces the fidelity by
an amount of the order of $(\Delta\bar{\vartheta})^2$
in the limit $\Delta\bar{\vartheta}\ll 1$.

The imprecision represented by $\Delta\Phi_j$ can be regarded as
an imprecision of the second $\pi$ pulse and hence is an~example of the case
which was treated in Sec.~\ref{imperfect}, namely
a~unitary error in a~$\pi$~pulse
having the~form of a rotation about the $z$ axis of the Bloch
sphere. The effect on the fidelity is given by Eq.~(\ref{Ftilde1})
with $\varepsilon_j=\Delta\Phi_j$ and $p_j=0$. For example,
in the case where $\Delta\Phi_j$ is caused by laser intensity
noise, the contribution to the infidelity is of order
$[C(\Delta I/I)]^2$, where $\Delta I$ is the fluctuation in laser
intensity between the two applications of the gate operation~$G$, 
and the constant~$C$ is of order $d / \xi a \simeq 500$. 
Thus, fluctuations of order $\Delta I/I\sim 10^{-3}$ 
would be sufficient to render the gate useless.

\subsection{Cancellation of single-qubit phase fluctuations}

We now propose a method to greatly reduce the sensitivity to
fluctuations.
We will assume that the fluctuations $\Delta \Phi_j$ are caused by
fluctuations in the pushing force, and we assume that the
force may vary, but $s$ does not. This case holds whenever
the force comes from a gradient of some potential, and the size but not
the shape of that potential can fluctuate. The analysis
therefore applies to the case of an optical dipole force with laser
intensity noise. An analysis with appropriate modifications would apply if
the source of fluctuations was another parameter such as the trapping
frequency $\omega$ hidden in the parameter $a$ or the trap distance $d$.

The essence of the idea is to work at a stationary
value of single-qubit phase, i.e. $d\Phi_j/d\xi=0$.
This can be done by choosing parameters such that the contribution from 
the light shift just cancels the contribution from the~Coulomb energy.
The Coulomb contribution (the~forth term in Eq.~(\ref{Phi1})) 
is proportional to the size of the force ($\xi\sim F$), 
while the light shift contribution (the~third term in Eq.~(\ref{Phi1})) 
is proportional to the size of light shift itself ($s\xi\sim E$).
Therefore, we have to balance the gradient
of the light shift with its size. This can be done by
choosing a convenient distance scale $s$ for the~laser beams.

To calculate the optimal value for $s$, we now have to reconsider
the analysis, taking force fluctuations into account.
The effect is that the parameter $\xi$ now depends on
time, where $\xi$ is expressed in Eq.~(\ref{force19}) giving
$\xi\propto|\Omega_0|^2\propto I$.
The analysis in previous sections is unchanged except that
wherever previously we wrote $\xi$, now we must put a mean
value $\l\xi\r$, where the~details of how the averaging takes place will
depend on how $\xi$ appears in the formulae, and the mean value is taken
over the~gate time $\tau$. We will not discuss such details,
but simply use the formulae, replacing $\xi^n$ by $\l\xi^n\r$.
To estimate the~changes in $\Phi_j$ due to fluctuations in $\xi$,
let us calculate the derivative of $\Phi_j$ giving
\be
 \label{dPhi}
\frac{d \Phi_1}{d \xi} =
\sqrt{ \frac{\pi}{8}\, } \epsilon \omega \tau
\left[
\frac{{\cal A}}{\epsilon}
-\frac{s}{a}\frac{\sqrt{8}}{\epsilon}
-\frac{d}{a}\frac{1}{\sqrt{2}}
+{\cal O}(a/d)
\right]\!,
\ee
where
\be
{\cal A}=
\left[
1 - \frac{\epsilon}{2} - \frac{1}{ (\omega\tau)^2}
\right]
\frac{d\l\xi^2\r}{d\xi}\,,
\ee
with a similar expression for $d \Phi_2/d \xi$
with the~replacements $s \rightarrow s'$ and $d \rightarrow -d$.

The pushing force $F$ is derived from a potential which we shall
call $V_F$, defined in such a way that for either ion, $x=0$ ($z=0$)
is the position occupied by the ion before the force is applied.
For example, for the travelling-wave (TW) and standing-wave (SW)
configuration discussed in Sec.~\ref{force} we may write
\be
\label{potential}
V_F=
\left\{
\begin{array}{lc}
V_0\,e^{-2[(x-x_0)/w]^2}, & \mbox{\ \ (TW)}\\[1mm]
V_0\sin^2[k(z-z_0)]\,, & \mbox{\ \ (SW)}
\end{array}
\right.
\ee
where $V_0=V_0(t)$ is a function of time.
In what follows we use the position variable $x$ but we understand that
it~changes to $z$ in the case of the standing-wave.
We introduced the potential energy into the Hamiltonian~(\ref{ion3})
in the form $(s-x\pm d/2)F$, where after the~transformation~(\ref{ion6}) 
it~is $(s-x)F$. Then, we have
\be
\label{poten}
(s-x)F\equiv V_F(x) = V_F(0) + x \frac{dV_F}{dx}(0) + {\cal O}(x^2)\,,
\nonumber\\
\ee
hence
\be
\label{sts}
s=-\frac{V_F(0)}{V_F'(0)}
=\left\{
\begin{array}{lc}
-w^2/4x_0 \,, & \mbox{\ \ (TW)} \\[1mm]
\tan(kz_0)/2k\,, & \mbox{\ \ (SW)}
\end{array}
\right.
\ee
where
$V_F'(0)=\frac{dV_F}{dx}(0)$ for TW, $V_F'(0)=\frac{dV_F}{dz}(0)$ for SW and
we neglected terms of order $x^2$ ($z^2$)
and above. The~same relations apply for
the parameter $s'$ (ion 2). We see from Eq.~(\ref{sts})
that $s$ and $s'$ can be tuned by adjusting
the~size of the beam waist $w$ (or standing-wave period $k$)
or the position $x_0$ ($z_0$) or both.
The reference position $x_0$ ($z_0$) and the waist $w$
can be chosen independently for each ion if so desired.
The standing-wave period $k$ is tunable without changing
the laser wavelength by adjusting the angle between the beams
forming the standing wave.

There exists a~``sweet spot" or an~optimal
choice for the~parameters $s$ and $s'$ which cancels $d \Phi_1/d\xi$ and
$d \Phi_2/d\xi$, namely
\bse
\label{sweet}
\be
s&=&
-\frac{\epsilon d}{4} + \frac{a}{\sqrt{8}}\,
\big[
{\cal A} + \epsilon\,{\cal O}(a/d)
\big]\,,\\[1mm]
s'&=&
+\frac{\epsilon d}{4} + \frac{a}{\sqrt{8}}\,
\big[
{\cal A} + \epsilon\,{\cal O}(a/d)
\big]\,,
\ee
\ese
which we obtained from Eq.~(\ref{dPhi}) for $d\Phi_j/d\xi=0$.
Typically, $\epsilon d\gg a$ so this implies that the parameters~$s$
and~$s'$ have opposite signs. There are two possibilities.
Either (i) the pushing forces act on the two ions
in the same direction, but the potential energy $V_F(0)$ has opposite
sign at the two ions, or (ii) the pushing forces act in opposite
directions and the potential energies have the same sign.

In the first case (forces in the same direction),
Eqs.~(\ref{sweet}) are valid.
However, either in the standing-wave configuration or
in the travelling-wave configuration (with both laser beams derived
from the same laser),
the~light shifts have the same sign at the two ions.
Therefore, we must have the second case (forces in opposite directions)
and we have to consider a situation different
to~the one treated up till now.
We could eventually consider the~first case if we applied 
opposite detunings for two consequent applications of the gate pulse~$G$. 
This would also directly implement the effect of the $\pi$-pulse method
without using any $\pi$ pulses \cite{paper2}.

With forces in opposite directions, a phase gate is still produced
with the same pulse timings, but now the~single-qubit
phases are changed. The simplest way to analyse this is to make
the~replacement $F_{\b}(t)\rightarrow -F_{\b}(t)$ in Eq.~(\ref{ion3}).
It means that we reverse the force on ion~2 (compared to ion~1)
but we preserve the relationships given
in Eq.~(\ref{ion4}) and (\ref{sts}). Carrying the analysis through,
we find that the single-particle potential energy phase is
\be
\phi'_{\b}=
\sqrt{\frac{\pi}{8}}\,\beta \omega\tau\xi^2+
\sqrt{\pi} \, \beta \omega \tau \xi \frac{s'}{a}\,,
\ee
in contrast this with Eq.~(\ref{ion27}). The two-particle phases
$\phi_{\a\b}$ are as described in Appendix~\ref{D}
with the replacement $\b\rightarrow -\b$.
Hence we now have
\be
\Phi_2=
\theta\left[
-\frac{1}{\epsilon(\omega\tau)^2}
+\frac{1}{\epsilon}
+\frac{s'}{a}\frac{\sqrt{8}}{\xi\epsilon}
-\frac{d}{a}\frac{1}{\xi\sqrt{2}}
-\frac{1}{2}
+{\cal O}(a/d)
\right],\nonumber\\
\ee
so the ``sweet spot" is
\be
\label{sweetp2}
s'&=&
+\frac{\epsilon d}{4}
-\frac{a}{\sqrt{8}}
\big[\,
{\cal A}+\epsilon\,{\cal O}(a/d)
\big]\,.
\ee
Note that $s$ and $s'$ still have opposite signs, but now
(\ref{sweet}a) and (\ref{sweetp2}) apply for the situation with
light shifts at the two ions of the same sign and
pushing forces on the two ions in opposite directions.
Example cases will be given in Sec.~\ref{infid}.

\section{Non-uniform pushing force}
\label{non-uni}

The original proposal \cite{ions2} assumed that the pushing force
is uniform. However, we introduced the optical dipole force as
a~suitable candidate for the pushing force, where the dipole force
originates from a non-uniform intensity profile of a laser beam.
Therefore, the ions experience a pushing force which depends on
their position. In their thermal vibrational motion the ions
explore different positions and as a result the force they
experience depends on their thermal motion. In this section we
will analyse this by allowing the dimensionless amplitude of the
force $\xi$ introduced in Eq.~(\ref{ion13}) to be a spatial
function rather than a constant (as we have considered up till
now).

Using $F(x,t)\,\,$=$\,\,-dV_F(x,t)/dx$ we obtain from Eq.~(\ref{potential}) 
the amplitude of the pushing force for the~travelling-wave configuration
\be
\label{non3}
\xi_{\rm trav}(x)=\xi_0\,\frac{x_0-x}{x_0}\,e^{-2(x^2-2x_0x)/w^2}\,,
\ee
where $\xi_0=\xi_{\rm trav}(0)$ and $x_0$ refers to the~ion position in
cross-section of the laser beam profile.
For the standing-wave configuration we get
\be
\label{non10}
\xi_{\rm stan}(z)=\xi_0\,\frac{\sin[2k(z_0-z)]}{\sin(2kz_0)}\,,
\ee
where $\xi_0=\xi_{\rm stan}(0)$ and $z_0$ refers to the~position of 
the~ion in the standing-wave field.

The non-uniformity of the pushing force introduces an additional
probabilistic character to the phase~$\Theta_{\a\b}$ because
the~amplitude~$\xi$ appears in all expressions for single-particle
and two-particle phases. Once the value of $\xi$ is not known
precisely it introduces a further imprecision to the system and,
consequently, a drop of the~fidelity.

The $\pi$-pulse method is again a useful tool because it goes
some way towards alleviating the effects of the
non-uniformity in the pushing force. In the full quantum approach
we would model a thermal state of
motion by a~mixture of vibrational energy eigenstates $\ket{n}$, then each
state $\ket{n}$ in the mixture acquires a different single-qubit
phase owing to the different average force it experiences. 
However, as long as
the heating rate is low, then the same phase will appear
in both implementations of the~operation~$G$ and so it is cancelled by 
the $\pi$-pulse method. Hence, recalling
Eq.~(\ref{err-opp}), (\ref{fid-perf}) and~(\ref{ion29}), we find that 
the~fidelity of the~phase gate with $\pi$~pulses will depend only on
two-particle phases.
To calculate the fidelity we need to estimate
the~difference $\delta\vartheta=\vartheta-\bar{\vartheta}$ between
the~actual and estimated value of the overall phase of the~gate.
In view of the~vibrational motion of the ion in the~non-uniform
laser beam profile, a good estimate for the~phases (compared to
Eq.~(\ref{ion32})) is now the average value
\be
\label{non0} 
\bar{\vartheta}= \int\limits_0^{2\pi}\frac{d\psi_1
d\psi_2}{(2\pi)^2} \int\limits_0^{+\infty}\frac{d\E_1
d\E_2}{(k_BT)^2}\,e^{-(\E_1+\E_2)/k_BT}\!\!
\int\limits_{-\infty}^{+\infty}dx\,\vartheta\,P(x)\,,\nonumber\\
\ee
where we model the ion's position by a Gaussian probability
distribution
\be
\label{non0.1}
P(x)=\frac{1}{\sigma\sqrt{2\pi}}\,e^{-x^2/2\sigma^2}.
\ee
The calculation method we have adopted here is slightly different
from the approach in Section~\ref{dp}. We have not analysed the ion
trajectories in detail. Owing to the~dependence of the force
on position they are now further complicated. However, in the limit
where the relative change in force with position is small, 
the ion samples a~small range of force values during its motion.
In the~adiabatic limit where many vibrational
oscillations occur during the gate time, we assume the main effect
is that the size of the force must be replaced by an average
value, but otherwise the previous calculation of the~dynamical phases
still applies. 

Eq.(\ref{non0.1}) is the positional distribution 
for a classical particle confined in a harmonic
potential at the temperature $T$ where 
the standard deviation $\sigma$ is given
by
\be
\label{non0.2}
\frac{1}{2}m\omega^2\sigma^2=\frac{1}{2}k_BT.
\ee
It is useful to express this in terms of the quantum harmonic
oscillator energy levels, giving
\be
\label{non0.3}
\sigma=\sqrt{\frac{k_BT}{m\omega^2}}\approx a\sqrt{\l n\r}\,,
\ee
where $\l n\r\approx k_BT/\hbar\omega$ is the mean vibrational number and
$a=\sqrt{\hbar/m\omega}$.

Now we have all the tools to calculate the fidelity
of the phase gate with $\pi$ pulses for the non-uniform pushing force.
The fidelity depends on the departures $\delta\vartheta$ of 
the~phase from its average (estimated) value, in other words on the variance of 
the~distribution of $\vartheta$ values. This is calculated in
Appendix~\ref{non-app}, where a complete expression, evaluated up to
order $(a/d)^4$ and to $(a/w)^4$ for the~travelling wave, and to all orders
in $a/\lambda$ for the~standing-wave case, is set out.
The most important terms in the~travelling-wave case are given by
\be
\label{non4}
{\cal F}'
&\simeq&
1-\left(\frac{6\theta_0k_BT}{\hbar\omega}\right)^2
\left(\frac{a}{d}\right)^4\nonumber\\[1mm]
& &
-\frac{2\theta_0}{3}\left(\frac{6\theta_0k_BT}{\hbar\omega}\right)
\left(\frac{a}{w}\right)^2
\left(
\frac{2x_0}{w}-\frac{w}{2x_0}
\right)^2\nonumber\\[1mm]
& &
-\frac{2}{9}\left(\frac{6\theta_0k_BT}{\hbar\omega}\right)^2
\left(\frac{a}{w}\right)^4
{\cal Q}(2x_0/w)\,,
\ee
where ${\cal Q}(y)=12y^4-64y^2+89-34/y^2+1/y^4$ and $2\theta_0\approx\pi$.
This expression has been written as a power series in $a/w$ and 
it is sufficiently accurate to give the behaviour for all the cases
considered in Sec.~\ref{infid}.
The first term corresponds to the fidelity with the uniform force 
in Eq.~(\ref{ion36}). However, in practice the other terms almost
always dominate because $w<d$.
The second term vanishes when $x_0=w/2$, but away
from this position it dominates. Therefore, placing the ion 
at $w_0=w/2$ offers a useful improvement in fidelity.

In the standing-wave case, the expression for the fidelity
simplifies considerably when the ion position is chosen as
$kz_0=\pi/4$. This is also the best choice to maximize 
the fidelity. Restricting to this choice for the~ion position,
the most important terms in the standing-wave case are 
\be
\label{non11}
{\cal F}'\simeq 
1-\frac{\theta_0^2}{32}
\left\{
1-\exp\left[-16(ka)^2\left(\frac{k_BT}{\hbar\omega}\right)\right]
\right\}^2.
\ee
This reproduces the full expression to good accuracy except
when ${\cal F}'\lesssim 0.9$.

\section{Total infidelity}
\label{infid}

We will now bring together all the contributions to infidelity
which have been discussed, in order to examine the~performance of
the gate in practice. These contributions are:
(i)~inaccuracy between actual $\Theta_{\a\b}$ and estimated phases 
$\bar{\Theta}_{\a\b}$, 
(ii)~imprecision of the~$\pi$~pulses, 
(iii)~photon scattering,
(iv)~laser intensity fluctuations and 
(v)~non-uniform character of the pushing force.
We will use parameters for the calcium ion to give examples.

Ignoring photon scattering for a moment, 
it follows from Eq.~(\ref{gc34}) that the fidelity of the phase gate is
\be
\label{infid1} 
\tilde{\cal F}=(1-4\zeta)(1-{\cal P}')\,, 
\ee 
where we introduced {\em infidelity} 
${\cal P'}\equiv 1-{\cal F'}$ and ${\cal F}'$ is given by Eq.~(\ref{non4})
or~(\ref{non11}), respectively. The effect of laser intensity 
fluctuations can be absorbed into the term $\zeta$ representing
imperfection of the $\pi$~pulses (see the~discussion in the last paragraph
in Sec.~\ref{fluct2}).

The influence of photon scattering further reduces 
the~fidelity. We have not calculated
in detail the combination of all effects at once, but the expected
behaviour is that the fidelities multiply, or the infidelities
add. In the limit of small infidelity these two statements agree.
Therefore, using Eq.~(\ref{infid3}) and~(\ref{infid1}),
we estimate the total infidelity, including all effects,
to be ${\cal F}_{\rm tot}={\cal F}'{\cal F}_{\rm scat}$, giving
\be 
\label{infid4} 
{\cal P}_{\rm tot}\simeq 4\zeta+{\cal P}'+ N\,,
\ee 
where we assumed all the contributions are small compared to one 
($\zeta, {\cal P}', N\ll 1$) and $2\theta_0\approx\pi$.

\subsection{Analysis}

We would like to minimize the infidelity while maximizing
desirable features such as gate speed. The overall problem has
11~parameters. Two are associated with the~design of the ion traps:
\bi
\item $\omega$ -- trapping frequency,
\item $d$ -- trap separation.
\ei
Four are associated with the laser beam:
\bi
\item $P$ -- laser power,
\item $w$ ($\pi/k$) -- size of the waist of the beam 
(period of the~standing wave),
\item $\Delta$ -- detuning of the laser,
\item $x_0$ ($z_0$) -- position of the ion in the profile of the~beam 
(in the standing-wave field).
\ei
Two describe unavoidable imperfections: 
\bi
\item $T$ -- temperature,
\item $\zeta$ -- imperfection of the $\pi$ pulses.
\ei
Finally, the last three parameters are associated
with the~ion and the transition which provides the light shift:
\bi
\item $m$ -- mass of the ion,
\item $\lambda_{\rm atom}$ -- wavelength of a dipole transition,
\item $\Gamma$ -- linewidth of that transition.
\ei
We will assume the last three parameters are ``given'',
i.e. they cannot be varied in seeking the best behaviour. We will
also assume that $\zeta$ is negligible ($\zeta\ll{\cal P}_{\rm tot}$). 
This implies that the $\pi$ pulses
are assumed to be precise, and that the laser intensity
noise is small. We find that in all cases to be considered, the force 
is only sufficient to displace the~ions by a small fraction of the trap
distance ($\bar{x}/d\sim 10^{-2}-10^{-3}$),
and therefore the ``sweet spot'' may
be needed to suppress the effect of intensity noise sufficiently
for this assumption to be valid.

The position of the ion in the laser beam should be
chosen sensibly. For the travelling wave, the temperature 
effects~(\ref{non4}) are minimized at $x_0=w/2$ 
(the~position of maximum force), but photon scattering~(\ref{force24}) 
is minimized at $x_0 =w/\sqrt{2}$. 
Since the fidelity~${\cal F}'$ depends more sharply on $x_0$ 
than the number of scattered photons, 
we will adopt the former choice, i.e. $x_0=w/2$. 
For the~standing wave we chose $kz_0=\pi/4$ (and $x_0=0$) 
for the~same reason.

The remaining parameters are 
$\omega$, $d$, $P$, $w$, $k$, $\Delta$ and $T$. We will
examine a range of values of $\omega$ and $d$, subject to
technological constraints on what is feasible for the trap
construction. For the temperature, we keep in mind that ion traps
in general have three temperature regimes which may be relevant:
(i)~the~Doppler limit temperature $k_BT=\hbar\Gamma/2$, 
(ii)~cooling to near the ground state 
$\l n\r\approx k_BT/\hbar\omega \simeq 1$, 
and (iii)~ground state cooling $\l n\r\ll 1$.
The regime~(i) is achieved by standard Doppler cooling techniques,
the regime~(iii) by the more technically demanding sideband cooling, and
the intermediate case is attractive because it can be achieved by
Doppler-type cooling on a narrow transition 
(i.e. using a~two-photon resonance). 
Since the gate under consideration is
relatively insensitive to temperature effects, we will avoid
assuming ground state cooling, by choosing either the Doppler
limit temperature or $\l n\r\simeq 1$.

Now it only remains to specify the laser parameters 
$P$, $w$, $k$ and $\Delta$. 
In most conditions higher laser power $P$ acts in a beneficial way,
so the correct choice is the highest power available (exceptions
to this rule can occur when the laser detuning $\Delta$ becomes of
the order of internal transition frequencies in the ion). 
The spot size~$w$ (for the~travelling wave) or the period~$\pi/k$ 
(for the~standing wave) has to make a compromise. A~smaller value increases
the force, which reduces the photon scattering, but the smaller
value also increases the sensitivity to temperature through
Eq.~(\ref{non4}) and (\ref{non11}).

It is noteworthy that the total fidelity does not depend on 
the laser detuning $\Delta$ (in the weak-coupling regime, i.e. 
$|\Omega|\ll|\Delta|$).
The detuning $\Delta$ can be adjusted to set 
the~size of the force ($\xi\propto 1/\Delta$) and hence
the~speed of the gate ($\tau\propto 1/\xi^2$).

\begin{figure}[htb]
\includegraphics[width=7.5cm]{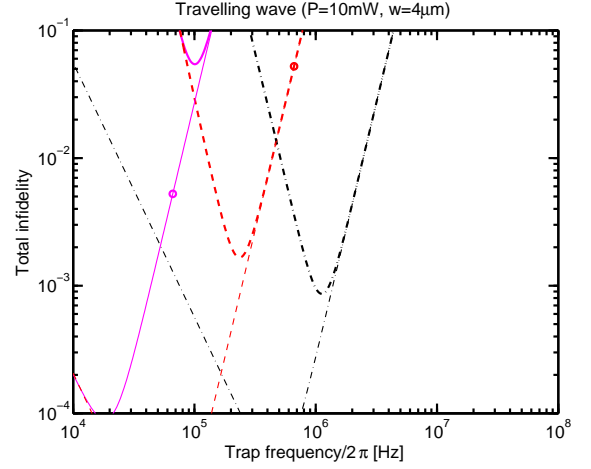}
\caption{The total infidelity given by Eq.~(\ref{infid4}) as
a~function of the trap frequency for various parameter
values for the travelling-wave configuration.
Results shown are for laser parameters $P=10\,{\rm mW}$
and $w=4\,\mu{\rm m}$. The solid lines are for the trap separation 
$d=100\,\mu$m, the~dashed lines are for $d=10\,\mu$m and the dash-dotted
lines are for $d=1\,\mu$m.
The thick lines are for the temperature equal to 
the~Doppler limit temperature associated with the dipole-allowed
$4S_{1/2}\leftrightarrow 4P_{1/2}$ transition at $\lambda=397\,{\rm nm}$
in Ca$^+$, i.e.~$T=538\,\mu{\rm K}$. 
The thin lines are for $k_BT=\hbar\omega$, i.e. $\l n \r\simeq 1$. 
The circles show the~``sweet spot'' condition~(\ref{omegasweet}).}
\label{trav_easy}
\end{figure}

\begin{figure}[htb]
\includegraphics[width=7.5cm]{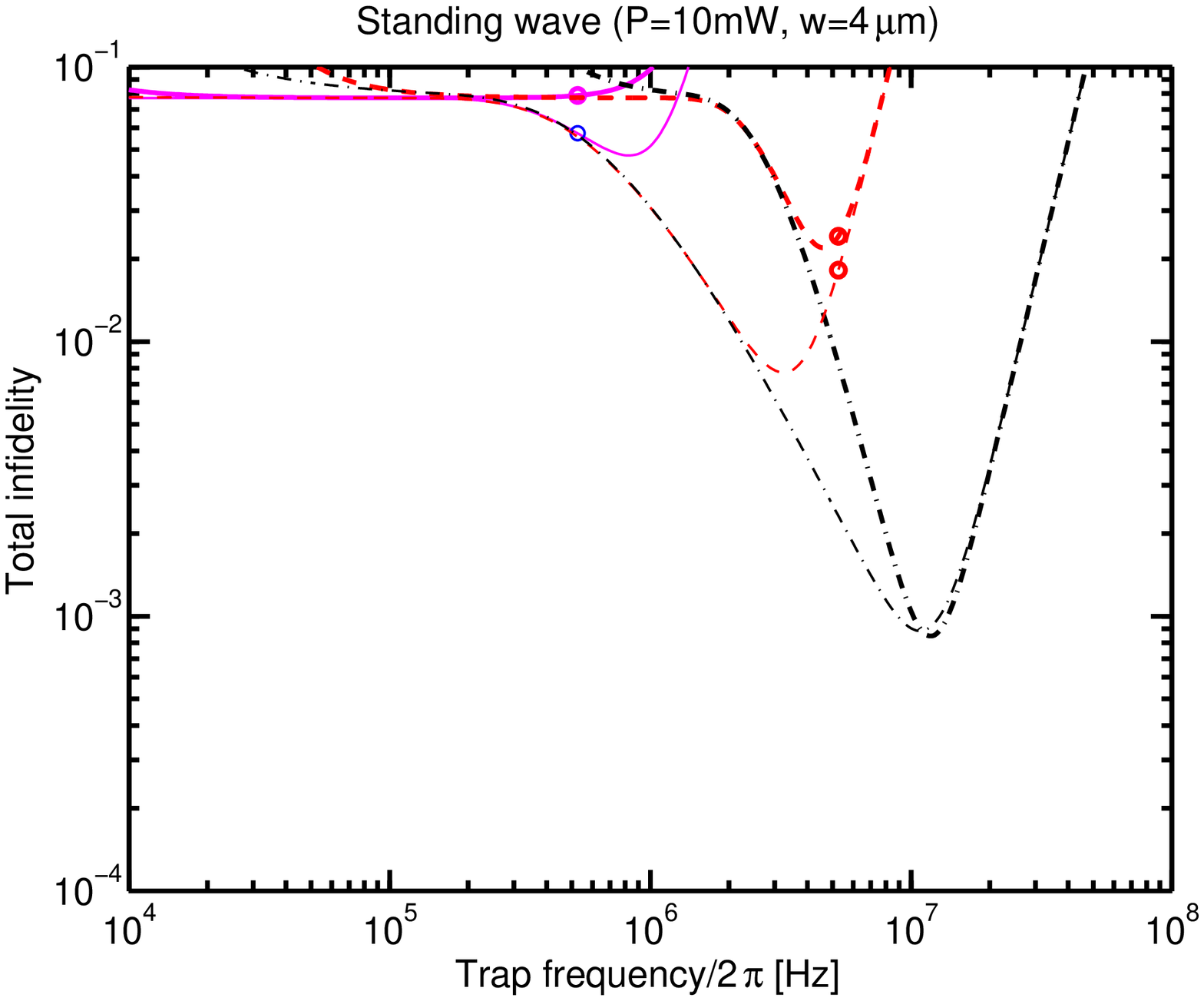}
\caption{Same as FIG.~\ref{trav_easy}, but for the standing-wave
configuration.}
\label{stan_easy}
\end{figure}

\begin{figure}[htb]
\includegraphics[width=7.5cm]{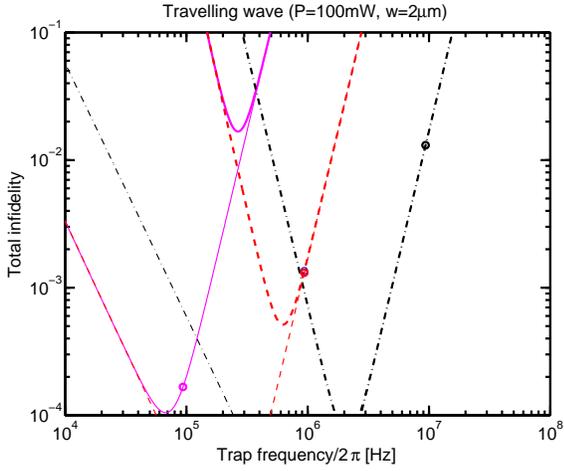}
\caption{Same as FIG.~\ref{trav_easy}, except now the laser parameters are
$P=100\,$mW and $w=2\,\mu$m.}
\label{trav_hard}
\end{figure}

\begin{figure}[htb]
\includegraphics[width=7.5cm]{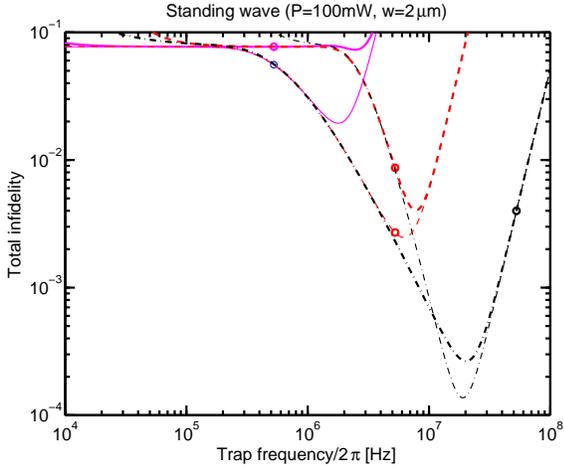}
\caption{Same as FIG.~\ref{trav_hard}, but for the standing-wave
configuration.}
\label{stan_hard}
\end{figure}

FIG.~\ref{trav_easy} and FIG.~\ref{stan_easy} 
show the total infidelity as a function of
trapping frequency, for various values of trap separation.
The~laser parameters are set equal to the~technically undemanding 
values $P=10\,{\rm mW}$ and $w=4\,\mu{\rm m}$. Results for two cooling 
regimes are shown:
either the~Doppler limit for calcium on a dipole-allowed
transition, or cooling to $\l n \r\simeq 1$. 
FIG.~\ref{trav_hard} and FIG.~\ref{stan_hard}
show results for the more demanding laser parameters
$P=100\,{\rm mW}$ and $w=2\,\mu{\rm m}$.
We see immediately that good fidelity can be attained, even at 
the Doppler limit temperature.

Next, let us examine $\bar{x}/d\sim a\xi/d$ in order to learn 
the~size of the single-qubit phases. 
Using Eq.~(\ref{ion16}) and (\ref{ion30}) we find
\be
\label{char}
\frac{a\xi}{d}\approx\sqrt{\frac{\omega d}{v}}\,,
\ee 
where we introduced the characteristic velocity 
$v=\omega\tau\ell\sqrt{8}/(\hbar\sqrt{\pi})\approx 1.6(\omega\tau)\a c$,
$\alpha$ is the fine structure constant, $c$ is the speed of light,
$\ell$ was defined in Eq.~(\ref{ion3}) and $2\theta\approx\pi$.
We would like $\bar{x}/d$ to be small but not too small. 
We~therefore choose the gate time $\tau$ to be small, which also
gives the fastest possible gate. The limit is set by the adiabatic
condition $\omega\tau\gg 1$. The non-adiabatic terms fall off
exponentially with $(\omega \tau)^2$ so a modest value 
$\omega\tau=5$ is sufficiently large, 
hence  $v=1.8\times 10^7\,{\rm m}\cdot{\rm s}^{-1}$.
FIG.~\ref{trav_easy}--\ref{stan_hard} show that the value of
$\omega d$ must be of order $10^2-10^3\,{\rm m}\cdot{\rm s}^{-1}$ 
in order to obtain high fidelity, therefore we find 
$a\xi/d$ is in the range $2\times 10^{-3}-8\times 10^{-3}$. 
This implies that in order to keep the single-qubit phase fluctuations 
sufficiently small it may be necessary to impose the ``sweet spot'' 
condition. For a given laser beam size and given trap separation, 
this is a condition on the~trapping frequency
\be
\label{omegasweet}
\omega_{\rm sweet}\approx\frac{1}{d}\sqrt{\frac{\ell}{m|s|}}\,,
\ee
where we approximated Eq.~(\ref{sweet}) to the form 
$|s|\approx\epsilon d/4$, since we find the parameter values are such 
that the second term in Eq.~(\ref{sweet}) is a small adjustment. 
The~``sweet spot'' values of $\omega$ are
indicated by the circles in FIG.~\ref{trav_easy}--\ref{stan_hard}.
If the ``sweet spot'' requirement is imposed, one can no longer tune
$\omega$ so as to minimize the infidelity.

A further constraint (in addition to the adiabatic approximation),
which affects the gate rate, is the condition that the ion is not pushed
too far (we will assume $\bar{x}\lesssim w/4$ for the~travelling wave 
and $\bar{x}\lesssim\lambda/10$ for the~standing wave).
Otherwise the ion will move into a~region where the light shift 
gradient is substantially reduced, the~net result would be a slower gate
and increased photon scattering. Using Eq.~(\ref{char}) we can write
\be
\label{taulimit}
\omega\tau\approx
\left(\frac{d}{\bar{x}_{\rm max}}\right)^2 
\frac{\omega d}{\alpha c}\,,   
\ee 
where $\bar{x}_{\rm max}=a\xi$ now denotes the maximum distance the~ion is
allowed to move. We choose $\bar{x}_{\rm max}=w/4$ 
($\bar{x}_{\rm max}=\lambda/10$) for
the~travelling (standing) wave. Taking the ``sweet spot''
value for $\omega d$ from Eq.~(\ref{omegasweet}), we find that
Eq.~(\ref{taulimit}) gives the limiting condition on $\omega\tau$ 
(i.e. $\omega\tau\gtrsim 5$)
for the~standing-wave case when $d\gtrsim 7\,\mu{\rm m}$ and 
for the~travelling-wave case when $d\gtrsim 220\,\mu{\rm m}$ 
at $w=2\,\mu{\rm m}$. 
In general one
might expect the~larger light shift gradient to allow the~gate to
be faster for the~standing wave compared to the~travelling wave.
Owing to the~previous findings, however this is no longer true when the~trap
separation $d$ is large, i.e. of the order of $50\,\mu{\rm m}$ or above.

We find the laser detuning $\Delta$ becomes comparable to the fine
structure splitting ($4P_{1/2}\leftrightarrow 4P_{3/2}$) of Ca$^+$ 
in the case of the travelling wave for 
$P=100\,{\rm mW}$ and $w=2\,\mu{\rm m}$, but 
in the~other cases the detuning is not too large. 
This would reduce the ratio of pushing force size 
to photon scattering rate, if the
qubit is stored in the ground state manifold, and therefore result
in more photon scattering.

We also find that $\epsilon\ll 1$ for the cases in 
FIG.~\ref{trav_easy}--\ref{stan_hard}, where the fidelity is high.

\subsection{Discussion}

Although our study has shown that the pushing gate in practice
does not offer as high a fidelity as was predicted in the initial
papers which described it, nevertheless comparing this gate with
others (which have been proposed for ion trap quantum computers),
it retains a~very good combination of allowed temperature and gate
rate. The 1995 proposal of Cirac and Zoller~\cite{CZ95} required
ground state cooling $\l n \r\ll 1$ and required 
the~gate rate~$R$ to be small compared to $\eta\omega$, 
where $\eta$ is the Lamb-Dicke parameter~\cite{steane00}. 
The work of M{\o}lmer and S{\o}rensen \cite{mol2,mol3, jon}
and others has revealed methods which allow 
the~requirement on cooling to be relaxed to $\l n\r\ll 1/\eta^2$. 
The ``pushing gate'', by contrast, allows a fast gate rate
$R=1/2\tau\simeq\omega/10$, with simultaneously 
high temperature. The standing-wave configuration places the stronger
constraint on temperature, but even for this case the Lamb-Dicke
condition $\l n \r\ll 1/\eta^2$ is sufficient.

The examples with $P=10\,{\rm mW}$ and $w=4\,\mu{\rm m}$ show that 
the gate could be demonstrated with modest requirements 
on the laser beam power and size. A trap separation below 
approximately $100\,\mu{\rm m}$ would be needed which is of the order 
of the smallest values which have
been reported for ion trap experiments up till now.

It is reasonable to expect that $d=10\,\mu{\rm m}$ may become available
in the (near) future. This would permit the~infidelity to fall to 
${\cal P}_{\rm tot}\sim 1\times 10^{-3}$ 
at the Doppler limit temperature, by using 
the travelling-wave configuration at $P=100\,{\rm mW}$, $w=2\,\mu{\rm m}$
and $\omega/2\pi=1\,{\rm MHz}$, giving a~gate time of 
$10/\omega=1.6\,\mu{\rm s}$. The same laser parameters 
with the standing-wave configuration would allow a gate time 
of $0.3\,\mu{\rm s}$ at infidelity ${\cal P}_{\rm tot}\sim 7\times 10^{-3}$.

In the longer term it would be desirable to have a~system
consisting of a large number ($>\!10^3$) of trap pairs, with many
gate operations performed simultaneously. With trap separations of
the order of micrometres and a~gate rate of order 10\,MHz, the infidelity of
order $4\times 10^{-4}$ would be available if the laser intensity
noise were small enough. However reducing the total laser power
may become an important consideration for a large processor.

\section{Conclusion}
\label{conc}

The main results of this paper are:

(i) The single-qubit rotations have contributions from the Coulomb
repulsion, the light (AC~Stark) shift, the~trapping potential,
and the kinetic energy. When the~parameter regime is such that
the two-qubit phase is insensitive to ion temperature, then the Coulomb
contribution to the single-qubit phases is much larger than $\pi$
($\sim 500\pi$) and proportional to the size of the pushing force.
This is dangerous because it implies that
laser intensity fluctuations just of the order of $10^{-3}$
could be sufficient to destroy the fidelity.

(ii) The dipole force was introduced. It is found that the number $N$
of scattered photons during the gate varies as $\lambda/P$ for
wavelength $\lambda$ and laser power $P$
for a laser beam focused such that the waist size is proportional
to the wavelength. Surprisingly,
photon scattering  is independent of the detuning $\Delta$ and the atomic
transition linewidth $\Gamma$ in the limit
$|\Omega|,\Gamma\ll|\Delta|$, where $\Omega$ is the~Rabi frequency.
The infidelity associated with photon scattering
increases with the trap separation $d$ and the trapping frequency
$\omega$, while the infidelity
associated with finite ion temperature falls with $d$ and $\omega$.
The best compromise is such that if the separation is
small, the~frequency should be large. It is found that
very good fidelity can be obtained for reasonable values of the~parameters
(see~FIG.~\ref{trav_easy}--\ref{stan_hard}).

(iii) A lack of spatial uniformity in the pushing force is found
to be a significant consideration when the distance scale ($w$ or $\pi/k$)
of the force profile is smaller than the separation $d$ of 
the ions. In this situation, the effect of a~finite temperature of
the ions is primarily to cause fluctuations in the size of 
the force they experience, and hence fluctuations in the phase of 
the~gate. Nevertheless, the gate performs well for reasonable values
of the~parameters, and cooling to the motional ground state is not
needed.

(iv) Using the $\pi$-pulse method (``spin-echo''),
we find that a given infidelity ($\zeta$ or $\varepsilon^2$)
in the $\pi$ pulses simply reduces the overall fidelity by approximately
$\zeta$ or $\varepsilon^2$.
An equivalent error occurs if a parameter of
the pushing gate changes slightly between two successive applications.
The most likely source of such fluctuations is laser intensity noise.
It is shown that the single-qubit phases can be made first order
insensitive to laser intensity noise, by choosing
the distance scale (``sweet spot'') of the laser intensity gradient so that
the AC~Stark light shift balances the contribution from Coulomb repulsion.
This greatly reduces the requirement on laser intensity
stability, indeed in practical terms it may make the difference between
a feasible gate and an unfeasible one.

To conclude overall, the analysis of the points summarized above
was necessary to assess the ``pushing gate'', and to gain insight into
its physical mechanism. The~analysis which has been presented is
sufficiently complete to describe an experimental realisation. 
Even though it~was found that the fidelity of the gate will not be as
high as was originally predicted, it still performs well in that
(i) it does not require ground state cooling and (ii) with sufficient
laser power neither is cooling to the Lamb-Dicke limit required,
while the gate time can be close to the inverse of trap vibrational
frequency, making it a fast gate in comparison with others which
have been proposed for ion traps.

\begin{acknowledgments}

We wish to acknowledge helpful conversations with T.~Calarco and
the opportunity to see some numerical results by V.~Barone and S.~Garelli.
This work was supported by the EPSRC and the Research Training,
Development and Human Potential Programs of the European Union 
(QUEST and QUBITS).

\end{acknowledgments}

\appendix
\section{Fidelity of diagonal unitary matrices}
\label{E}

Suppose a physical operation designed to produce some logical operation
$U$ instead produces $G$, where $G\neq U$. The fidelity is then defined to be
\be
\label{f1}
{\cal F}=
\min_{|\Psi_0\r} | \bra{\Psi_0} U^{\dagger} G \ket{\Psi_0} |^2\,,
\ee
where
\be
\label{f2}
|\Psi_0\r=\sum_{\a,\b=0}^1 c_{\a\b} |\a\b\r
\ee
is an arbitrary two-qubit state denoting $|\a\b\r=|\a\r\otimes |\b\r$.

Next we will calculate
${\cal F}$ for some cases of interest when the operator $U^{\dagger} G$ is
diagonal and unitary. Therefore, let us consider
a general diagonal operator of two qubits
\be
\label{f3}
U^{\dag}G={\rm diag}
\{
e^{i\Theta_{00}}, e^{i\Theta_{01}}, e^{i\Theta_{10}}, e^{i\Theta_{11}}
\}\,.
\ee
Without a loss of generality we may extract $\Theta_{00}$ as a~global phase
and relabel Eq.~(\ref{f3}) as follows
\be
\label{f4}
U^{\dag}G={\rm diag}
\{
1, e^{i\theta_{01}}, e^{i\theta_{10}}, e^{i\theta_{11}}
\}\,.
\ee
When we now substitute Eq.~(\ref{f4}) to (\ref{f1}) we get
\be
\label{f5}
{\cal F}=\min_{w,x,y,z}(1-4f)
\ee
with
\be
\label{f6}
f
&=&
wx\,\sin^2(\theta_{01}/2)
+wy\,\sin^2(\theta_{10}/2)
+wz\,\sin^2(\theta_{11}/2)\nonumber\\[1mm]
& &
+xy\,\sin^2\big[(\theta_{01}-\theta_{10})/2\big]
+xz\,\sin^2\big[(\theta_{01}-\theta_{11})/2\big]\nonumber\\[1mm]
& &
+yz\,\sin^2\big[(\theta_{10}-\theta_{11})/2\big]\,,
\ee
where to make the expressions less cluttered we used the~notation
\be
\label{f7}
\{w,x,y,z \}
\equiv
\{ |c_{00}|^2,|c_{01}|^2,|c_{10}|^2,|c_{11}|^2 \}\,.
\ee
The normalisation $w+x+y+z=1$ may be used to reduce $f$
to a function of three variables $f=f(x,y,z)$ with three parameters
$\theta_{01}, \theta_{10}, \theta_{11}$.

To find the minimum fidelity ${\cal F}$, the function $f$ must be maximised.
The general case is straightforward to calculate but leads
to a~complicated expression
without much useful insight. It is more valuable to consider some
special cases.

(i) When one phase is negligible compared to the others
(for instance $\theta_{11}\ll\theta_{01},\theta_{10}$),
using the normalisation condition we have
\be
\label{f8}
f(x,y)
&\approx&
(1-x-y)
\big[
x\sin^2(\theta_{01}/2)+y\sin^2(\theta_{10}/2)
\big]\nonumber\\[1mm]
& &
+xy\sin^2\big[(\theta_{01}-\theta_{10})/2\big].
\ee
Since this is now a function of only two variables $x$ and $y$ the
maximisation is easier to calculate but still does not yield a
concise answer.

(ii) For the case $\theta\equiv\theta_{10}=\theta_{01}\gg\theta_{11}$,
from Eq.~(\ref{f8}) we have
\be
f(x,y)=
(1-x-y)(x+y)\sin^2(\theta/2)\,,
\ee
whose maximum value is $f_{\rm max}=(1/4)\sin^2(\theta/2)$
when $x+y=1/2$. Hence the fidelity is
\be
\label{f10}
{\cal F} = 1-\sin^2(\theta/2) = \cos^2(\theta/2)\,.
\ee

(iii) For the case  $\theta\equiv\theta_{10}=-\theta_{01}\gg\theta_{11}$,
from Eq.~(\ref{f8}) we have
\be
f(x,y)=
(1-x-y)(x+y)\sin^2(\theta/2) + xy\sin^2{\theta}\,.\nonumber\\
\ee
The maximum value is on the line $x=y$, where after some algebra we get from
the~previous expression
\be
f(x)= 2x\sin^2(\theta/2) \big[1 - 2x\sin^2(\theta/2)\big].      \label{f12}
\ee
This function has a~maximum value $f_{\rm max}=1/4$ when $\theta\ge\pi/2$,
yielding zero fidelity (${\cal F}=0$).
For $\theta\le\pi/2$ the~maximum of the~function~$f$
is when $x$ takes its largest value subject to the normalisation
constraint $2x\leq 1$, giving $f_{\rm max}=(1/4)\sin^2{\theta}$.
Hence the fidelity is
\be
{\cal F}=\cos^2\theta        \label{fct}
\ee

(iv) Finally, if all phases apart from one are negligible
(for instance $\theta\equiv\theta_{01}\gg\theta_{10},\theta_{11}$),
we have
\be
\label{f13}
f(x)=
x(1-x)\sin^2(\theta/2)\,,
\ee
giving the fidelity ${\cal F}=\cos^2(\theta/2)$.

\section{Analysis of pure ``over-rotation'' errors of $\pi$~pulses}
\label{F}

To study the effect of ``over-rotation'' in the~$\pi$~pulses, 
we assume that a single-qubit gate intended to be a~$\pi$~pulse on ion~$j$
in fact produces the effect $R_j M_j(\varepsilon_j)$, where 
$M$ is given in Eq.~(\ref{gc25}).
Therefore, the complete gate sequence replaces the one in Eq.~(\ref{gc20}),
giving $S'(RM_{\rm B})G(RM_{\rm A})G$,
where 
\bse
\be
M_{\rm A}&=&M_1(\varepsilon_1)\otimes M_2(\varepsilon_2)\,,\\
M_{\rm B}&=&M_1(p\varepsilon_1)\otimes M_2(p \varepsilon_2)\,.
\ee
\ese
Recalling Eq.~(\ref{fprime}), we find the fidelity is
$\min\left|\l\Psi_0|Q|\Psi_0\r\right|^2$, where
\be
Q 
&=& 
\left[  
(G_{\rm perf}^{\dag}R^{\dag})^2 (S^{\p})^{\dag} 
\right]
\left[ 
S'(RM_{\rm B})G(RM_{\rm A})G 
\right]\nonumber\\[1mm]
&=&  
(G_{\rm perf}^{\dag}R^{\dag})G_{\rm perf}^{\dag}M_{\rm B}G(RM_{\rm A})G\,.
\ee
This operator, and the fidelity associated with it, does not
readily simplify. We studied it by a combination of numerical
searches for least fidelity, and by algebraic treatment of some less general
cases. It was found that the case where both ions have the same
``over-rotation'' ($\varepsilon_1=\varepsilon_2$) does not lead to
significantly different results from the case where the ions have different
``over-rotation''. In this slightly restricted case, and considering
for simplicity a perfect phase gate ($G=G_{\rm perf}$), we obtain
\be
Q 
&\approx& 
\openone+\frac{i\varepsilon}{2} 
\left[ 
\begin{array}{cccc}
0 & p+i & p+i & 0 \\
p-i & 0 & 0 & -(1+p) \\
p-i & 0 & 0 & -(1+p) \\
0\ & -(1+p)\ & -(1+p)\ & 0       
\end{array} 
\right]  
\nonumber\\[2mm]
& & 
-\frac{\varepsilon^2}{4} 
\left[ 
\begin{array}{cccc}
1+p^2 - 2ip\ & 0\ & 0\ & -(2p+p^2+i) \\
0\ & \chi\ & \chi\ & 0 \\
0\ & \chi\ & \chi\ & 0 \\
-p^2 + i(1+2p)\ & 0\ & 0\ & (1+p)^2 
\end{array} 
\right],
\nonumber
\ee
where $\chi= 1+p^2+(1+i)p$.
The fidelity obtained from this is of the form 
$(a+bp+cp^2)\epsilon^2$, where the coefficients $a,b,c$ depend on 
the~initial state~$|\Psi_0\r$ under consideration. 
The~main properties of this result are that the fidelity is
found to be proportional to $\varepsilon^2$ (not $\varepsilon$)
and there is no special case when $p=1$ or $p=-1$. That means 
the~effects of two ``over-rotations'' by the same or by opposite 
amounts do not cancel.

\section{Fidelity of the phase gate with imperfect $\pi$~pulses}
\label{A}

The single-qubit rotations acting on both qubits expressed by
Eq.~(\ref{gc4})--(\ref{gc5}) can be written (up to global phases) as
\be
\label{a1}
S
&=&
\sum_{\a,\b=0}^1\,|\a\b\r\l\a\b|\,e^{-i\bar{\Theta}_{\a\b}}
\,\big(e^{i\bar{\vartheta}}\big)^{\a\b}\\
&=&
|00 \r\l 00|\,e^{-i\bar{\Theta}_{00}}+
|01 \r\l 01|\,e^{-i\bar{\Theta}_{01}}\nonumber\\[1mm]
& &
+|10 \r\l 10|\,e^{-i\bar{\Theta}_{10}}+
|11 \r\l 11|\,e^{-i\bar{\Theta}_{11}} e^{i\bar{\vartheta}}\,.\nonumber
\ee
Similarly, we can rewrite the rotations $\tilde{S}$
in Eq.~(\ref{gc30})--(\ref{gc31}) to the form
\be
\label{a2}
\tilde{S}
&=&
\sum_{\a,\b=0}^1\,|\a\b\r\l\a\b|\,e^{-i\bar{\Theta}_{\a\b}}
\,\big(e^{i\bar{\vartheta}}\big)^{\delta_{\a\b}-\a\b}\\
&=&
|00 \r\l 00|\,e^{-i\bar{\Theta}_{00}} e^{i\bar{\vartheta}}+
|01 \r\l 01|\,e^{-i\bar{\Theta}_{01}}\nonumber\\[1mm]
& &
+|10 \r\l 10|\,e^{-i\bar{\Theta}_{10}}+
|11 \r\l 11|\,e^{-i\bar{\Theta}_{11}}\,,\nonumber
\ee
where $\delta_{\a\b}$ is the Kronecker delta symbol.
The two different single-qubit operations
$S$ and $\tilde{S}$ are used here, rather than a single operation
at the end, because it may be advantageous to run an experiment this way, 
since then imperfections in $G$ can be undone immediately,
using a feed-forward technique on the hardware which implements
the gates.

Even though we introduced different bit-flip error probabilities
$\zeta_1$ and $\zeta_2$ for the qubit 1 and 2 in Eq.~(\ref{gc27}),
they will be similar in practice $(\zeta_1\simeq\zeta_2)$.
Here we will assume
$\zeta_1=\zeta_2=\zeta$, where the error is very small ($\zeta\ll 1$).
We note that the overall fidelity of the whole sequence is similar
in the~cases $\zeta_1=\zeta_2$ and $\zeta_1\simeq\zeta_2$.

Following the complete gate sequence (\ref{gc29})
we can now calculate the actual state of the system
(representing the~evolution of the phase gate with imperfect $\pi$ pulses)
using Eq.~(\ref{a1}) and (\ref{a2}), that is
\be
\label{a3}
\rho_{\rm act}
&\simeq&
(1-4\zeta)\,|\,\Xi_0\r\l\Xi_0|\nonumber\\
& &
+\zeta\sum_{j=1}^4|\,\Xi_j\r\l\Xi_j|+{\cal O}(\zeta^2)\,,
\ee
where
\begin{eqnarray*}
|\,\Xi_0\r&=&
\sum_{\a,\b}\,
c_{\a\b}\,e^{i(\Theta_{\a\b}+\Theta_{\ap\bp})}\,
\big(e^{i\bar{\vartheta}}\big)^{\a\b+\delta_{\ap\bp}-\ap\bp}\,|\a\b\r\,,\\
|\,\Xi_1\r&=&
\sum_{\a,\b}\,
c_{\a\b}\,e^{i(\Theta_{\a\b}+\Theta_{\ap\bp})}\,
\big(e^{i\bar{\vartheta}}\big)^{\a\b+\delta_{\ap\bp}-\ap\bp}\,|\a\bp\r\,,\\
|\,\Xi_2\r&=&
\sum_{\a,\b}\,
c_{\a\b}\,e^{i(\Theta_{\a\b}+\Theta_{\ap\bp})}\,
\big(e^{i\bar{\vartheta}}\big)^{\a\b+\delta_{\ap\bp}-\ap\bp}\,|\ap\b\r\,,\\
|\,\Xi_3\r&=&
\sum_{\a,\b}\,
c_{\a\b}\,e^{i(\Theta_{\a\b}+\Theta_{\ap\b})}\,
\big(e^{i\bar{\vartheta}}\big)^{\a\b+\delta_{\ap\b}-\ap\b}\,|\a\bp\r\,,\\
|\,\Xi_4\r&=&
\sum_{\a,\b}\,
c_{\a\b}\,e^{i(\Theta_{\a\b}+\Theta_{\a\bp})}\,
\big(e^{i\bar{\vartheta}}\big)^{\a\b+\delta_{\a\bp}-\a\bp}\,|\ap\b\r\,.
\end{eqnarray*}
All the summations run over $\a,\b=0,1$,
we used the~notation $\ap\equiv 1-\a$, $\bp\equiv 1-\b$ and
applied the~approximation
\be
\label{a4}
(1+\zeta)^{m}\approx 1+m\zeta\,,\qquad \zeta\ll 1\,.
\ee
In what follows we neglect terms of the second and higher
orders in $\zeta$ denoted as ${\cal O}(\zeta^2)$.

Substituting Eq.~(\ref{a3}) into Eq.~(\ref{gc33})
the fidelity of the~phase gate with the~imperfect $\pi$~pulses is
\be
\label{b1}
\tilde{\cal F}
&\simeq&
\bigg\l
\min\limits_{\{c_{\a\b}\}}
\bigg\{
(1-4\zeta)
\bigg|
\sum_{\a,\b}|c_{\a\b}|^2e^{i(\delta\Theta_{\a\b}+\delta\Theta_{\ap\bp})}
\bigg|^2
\nonumber
\ee
\vspace{-6mm}
\be
&+&
\hspace{-2mm}
\zeta\left|
({K}\!-\!{N})\,e^{i(\delta\Theta_{00}+\delta\Theta_{11})}+
({K}^*\!-\!{N}^*)\,e^{i(\delta\Theta_{01}+\delta\Theta_{10})}
\right|^2\nonumber\\[1mm]
&+&
\hspace{-2mm}
\zeta\left|
({L}\!-\!{M})\,e^{i(\delta\Theta_{00}+\delta\Theta_{11})}+
({L}^*\!-\!{M}^*)\,e^{i(\delta\Theta_{01}+\delta\Theta_{10})}
\right|^2\nonumber\\[1mm]
&+&
\hspace{-2mm}
\zeta\left|
({K}\!-\!i{N}^*)\,e^{i(\delta\Theta_{00}+\delta\Theta_{10})}+
({K}^*\!+\!i{N})\,e^{i(\delta\Theta_{01}+\delta\Theta_{11})}
\right|^2\nonumber\\[1mm]
&+&
\hspace{-2mm}
\zeta\left|
(L\!-\!iM^*)\,e^{i(\delta\Theta_{00}+\delta\Theta_{01})}+
(L^*\!+\!iM)\,e^{i(\delta\Theta_{10}+\delta\Theta_{11})}
\right|^2\!\bigg\}\!\bigg\r\nonumber\hspace*{-2mm}\\
\ee
where
$\{K,L,M,N\}\equiv
\{
c_{00}\,c_{01}^*,\,
c_{00}\,c_{10}^*,\,
c_{11}\,c_{01}^*,\,
c_{11}\,c_{10}^*
\}
$ and $\a,\b=0,1$.
The first term in Eq.~(\ref{b1}) is equal (up to the factor~$1-4\zeta$) to
the~fidelity of the gate with the~perfect $\pi$~pulses expressed by
Eq.~(\ref{gc24}). The other four terms in Eq.~(\ref{b1}) have always
a~non-negative contribution. We can find an initial state
(i.e. a~combination of coefficients $c_{\a\b}$) which makes the contribution
of these four terms equal to zero, thus minimising
the~expression in Eq.~(\ref{b1}). Then the fidelity of the phase gate with
imperfect $\pi$~pulses is as given in Eq.~(\ref{gc34}).

\section{Forced classical harmonic oscillator}
\label{C}

We treat the ion system semiclassically, in that we assume internal
ionic states $|0\r$ and $|1\r$ but we treat the motion of the ions as 
a~motion of classical particles. Then, the motion of a single classical
particle confined in a harmonic potential
with an external force is described by the~Lagrangian
\be
\label{c1}
L(t)=
\frac{1}{2}m\dot{x}^2-\left[\frac{1}{2}m\omega^2x^2-xF(t)\right]\,,
\ee
where $x(t)$ denotes the trajectory of the particle,
$F(t)$ is a time-dependent external force and $\dot{x}=dx/dt$.
We can rewrite the~Lagrangian~(\ref{c1}) in the form
\be
\label{c2}
L(t)=
\frac{1}{2}m\dot{x}^2-\frac{1}{2}m\omega^2(x-\bar{x})^2-
\frac{1}{2}m\omega^2\bar{x}^2\,,
\ee
where
\be
\label{c3}
\bar{x}(t)=F(t)/m\omega^2\,.
\ee
It follows from Eq.~(\ref{c1}) and (\ref{c2}) that the action of the~force
on the particle corresponds to the~time-dependent displacement~(\ref{c3})
of the harmonic potential in which the particle is confined. The dynamics of
the~particle are given by the Lagrange equation
\be
\label{c4}
\frac{d}{dt}\frac{\partial L}{\partial\dot{x}}-
\frac{\partial L}{\partial x}=0\,.
\ee
Substituting the Lagrangian~(\ref{c2}) into Eq.~(\ref{c4}) we get
the~equation of motion
\be
\label{c5}
\ddot{x}(t)+\omega^2x(t)=F(t)/m\,,
\ee
which can be solved analytically. The solution splits into three terms
\be
\label{c6}
x(t)=\bar{x}(t)-\delta (t)+\Delta (t)\,,
\ee
where $\bar{x}(t)$ is the time-dependent displacement of the potential
defined in Eq.~(\ref{c3}), $\delta (t)$ is so called {\it sloshing motion}
\be
\label{c7}
\delta(t)
&=&
\sin(\omega t)\left[
\int_{t_0}^t\dot{\bar{x}}(t^{\p})\sin(\omega t^{\p})\,dt^{\p}
\right]\nonumber\\[1mm]
& &
+\cos(\omega t)
\left[\int_{t_0}^t\dot{\bar{x}}(t^{\p})\cos(\omega t^{\p})\,dt^{\p}\right],
\ee
and finally $\Delta (t)$ denotes standard oscillations of the particle
confined in the harmonic potential
\be
\label{c8}
\Delta(t)=
(x_0-\bar{x}_0)\cos(\omega t)
+(\dot{x}_0/\omega)\sin(\omega t)\,,
\ee
where the initial conditions are
$x_0=x(t_0)$, $\bar{x}_0=\bar{x}(t_0)$ and 
$\dot{x}_0=\frac{dx}{dt}(t_0)$.
We can rewrite Eq.~(\ref{c8}) in the form
\be
\label{c9}
\Delta (t)=\sqrt{2\E/m\omega^2} \; \cos(\omega t+\psi)\,,
\ee
where $\E$ is the oscillation energy
of the particle in the~harmonic potential and
$\psi$ is the initial phase which depends on initial conditions.

We can assume that the ion experiences a harmonic potential only in 
the~limit $\epsilon\ll 1$. Otherwise, Eq.~(\ref{c6}) does not describe its
motion or describes it only approximately.

\section{Two-particle interaction phases}
\label{D}

It is sufficient to consider in Eq.~(\ref{ion28})
with respect to assumptions~(i)--(iii)
only the terms up to $n=4$ and we choose
$|t_0|,|t|\to\infty$ (except for global phases)
for the~same reason as it is stated in Eq.~(\ref{ion24}).
Then we can have (swapping integration and summation)
\be
\label{d1}
\phi_{\a\b}\approx
\phi_{\a\b}^{(0)}
+\sum_{n=1}^4\phi_{\a\b}^{(n)}\,,
\ee
where
\be
\label{d2}
\phi_{\a\b}^{(n)}=
-\frac{\ell}{\hbar d}\int\limits_{-\infty}^{+\infty}
\left[\frac{x_{\a}(t)-x_{\b}^{\p}(t)}{d}\right]^ndt\,,
\ee
and $\phi_{\a\b}^{(0)}=-(\ell/\hbar d)(t-t_0)$ is a global phase.
Using Eq.~(\ref{ion15}) the~motion of ion~1
in the~internal state $|\a\r$ and ion~2 in the state $|\b\r$ can be
expressed (when $\epsilon\ll 1$) as follows
\bse
\label{d3}
\be
x_{\a}(t)&=&\bar{x}_{\a}(t)+\Delta(t)\,,\\
x_{\b}^{\p}(t)&=&\bar{x}_{\b}^{\p}(t)+\Delta^{\p}(t)\,,
\ee
\ese
where the time-dependent displacements $x_{\a}(t)$ and $x_{\b}^{\p}(t)$ are
defined by Eq.~(\ref{ion9}) while the oscillations in the~microtraps
(Eq.~(\ref{c9})) are
\bse
\label{d4}
\be
\Delta(t)&=&\sqrt{\frac{2\E_1}{m\omega^2}}\,\cos(\omega t+\psi_1)\,,\\
\Delta^{\p}(t)&=&\sqrt{\frac{2\E_2}{m\omega^2}}\,\cos(\omega t+\psi_2)\,.
\ee
\ese
The parameters
$\E_1$ and $\E_2$ are oscillations energies of the~ions which are in
principle different because the ions are trapped in two separate and
independent potentials. Using Eq.~(\ref{d3}) and~(\ref{d4}) we can express
Eq.~(\ref{d2}) in the~form
\be
\label{ddd}
\phi_{\a\b}^{(n)}=
-\frac{\ell}{\hbar d}\int\limits_{-\infty}^{+\infty}
\left[\frac{(\a-\b)a{\cal F}+(\Delta-\Delta^{\p})}{d}\right]^n dt\,.
\ee
When we now neglect in the adiabatic approximation~(\ref{ion14})
all terms where the factor $e^{-(\omega\tau)^2}$ appears, then
we can express the linear term ($n=1$) from Eq.~(\ref{ddd})
as follows
\be
\label{d5}
\phi_{\a\b}^{(1)}=
-\frac{\sqrt{\pi}}{4}\,(\a-\b)\epsilon\omega\tau\xi
\,\frac{d}{a}+{\rm Q}_1\,,
\ee
where $\a,\b=0,1$, $\epsilon$ is defined by Eq.~(\ref{ion16}),
$\xi$ and $\tau$ are
introduced in Eq.~(\ref{ion13}), $\omega$ is the oscillation frequency with
respect to Eq.~(\ref{ion17}), $d$ is the ion separation with respect to
Eq.~(\ref{ion19}), $a$ is defined in Eq.~(\ref{ion5}) and
for global phases we used the notation
\be
\label{d6}
{\rm Q}_n=-\frac{\ell}{\hbar d}\int_{t_0}^{t}
\left(\frac{\Delta-\Delta^{\p}}{d}\right)^n dt^{\p}\,.
\ee
The quadratic term ($n=2$) reads
\be
\label{d7}
\phi_{\a\b}^{(2)}=
-\sqrt{\frac{\pi}{32}}\,(\a-\b)^2\epsilon\omega\tau\xi^2+{\rm Q}_2\,,
\ee
the cubic term ($n=3$) reads
\be
\label{d8}
\phi_{\a\b}^{(3)}
&=&
-\sqrt{\frac{\pi}{48}}\,(\a-\b)^3\epsilon\omega\tau\xi^3
\,\frac{a}{d}\nonumber\\[1mm]
& &
-\frac{3}{4}\sqrt{\pi}\,(\a-\b)\epsilon\omega\tau\xi
\,\frac{a}{d}\times\nonumber\\[2mm]
& &
\times\left[\frac{\E_1}{\hbar\omega}+\frac{\E_2}{\hbar\omega}
-2\frac{\sqrt{\E_1\E_2}}{\hbar\omega}\cos(\psi_1-\psi_2)\right]
+{\rm Q}_3\nonumber\\
\ee
and finally the biquadratic term ($n=4$) reads
\be
\label{d9}
\phi_{\a\b}^{(4)}
&=&
-\sqrt{\frac{\pi}{8}}\,(\a-\b)^4\epsilon\omega\tau\xi^4
\left(\frac{a}{d}\right)^2\nonumber\\[1mm]
& &
-\sqrt{\frac{9\pi}{8}}\,(\a-\b)^2\epsilon\omega\tau\xi^2
\left(\frac{a}{d}\right)^2\times\nonumber\\[2mm]
& &
\times\left[\frac{\E_1}{\hbar\omega}+\frac{\E_2}{\hbar\omega}
-2\frac{\sqrt{\E_1\E_2}}{\hbar\omega}\cos(\psi_1-\psi_2)\right]
+{\rm Q}_4\,,\nonumber\\[1mm]
\ee
where $\E_1,\,\E_2$ and $\psi_1,\,\psi_2$ are associated with oscillations of
the~ions in the microtraps and they are defined in Eq.~(\ref{d4}).

When we substitute Eq.~(\ref{ion30}) into Eq.~(\ref{d5})--(\ref{d9})
we can write the interaction phases $\phi_{\a\b}$ in Eq.~(\ref{ion23})
with the~precision up to $n=4$ in a~compact form
\be
\label{d10}
\phi_{\a\b}
&=&
-(\a-\b)\,\theta\,\frac{d}{a}\frac{1}{\xi\sqrt{2}}
-(\a-\b)^2\theta\frac{1}{2}
\nonumber\\[1mm]
& &
-(\a-\b)^3\,\theta\,\frac{a}{d}
\left\{\frac{\xi}{\sqrt{6}}\right.+\frac{3}{\xi\sqrt{2}}\times
\nonumber\\[1mm]
& &
\times
\left.
\left[
\frac{\E_1}{\hbar\omega}+\frac{\E_2}{\hbar\omega}
-2\frac{\sqrt{\E_1\E_2}}{\hbar\omega}\cos(\psi_1-\psi_2)
\right]
\right\}
\nonumber\\[1mm]
& &
-(\a-\b)^4\,\theta\left(\frac{a}{d}\right)^2
\left\{\frac{\xi^2}{\sqrt{8}}\right.+
\nonumber\\[1mm]
& &
\left.
+3\left[
\frac{\E_1}{\hbar\omega}+\frac{\E_2}{\hbar\omega}
-2\frac{\sqrt{\E_1\E_2}}{\hbar\omega}\cos(\psi_1-\psi_2)
\right]
\right\}.
\nonumber\\[1mm]
\ee
The mean (estimated) value of the phases $\phi_{\a\b}$ 
with respect to Eq.~(\ref{ion32})
is
\be
\label{d11}
\bar{\phi}_{\a\b}
&=&
-(\a-\b)\,\theta\,\frac{d}{a}\frac{1}{\xi\sqrt{2}}
-(\a-\b)^2\,\theta\,\frac{1}{2}
\nonumber\\[1mm]
& &
-(\a-\b)^3\,\theta\,\frac{a}{d}
\left(
\frac{\xi}{\sqrt{6}}+\frac{1}{\xi\sqrt{2}}\frac{6k_BT}{\hbar\omega}
\right)
\nonumber\\[1mm]
& &
-(\a-\b)^4\,\theta\left(\frac{a}{d}\right)^2
\left(
\frac{\xi^2}{\sqrt{8}}+\frac{6k_BT}{\hbar\omega}
\right).
\ee
Now we can easily calculate the difference between the~actual and 
the~estimated value of the phase $\phi_{\a\b}$ which appears 
in the expression for
the fidelity of the gate in Eq.~(\ref{gc15}) and (\ref{gc24}).
We will find out (in the approximation when we ignore fluctuations
of the pushing force) that the fidelity of the gate is determined by 
the~terms of the order $n=3$ and $n=4$ which represent
thermal corrections to the motion of the ions.

\section{Non-uniform pushing force: Fidelity}
\label{non-app}

It follows from Eq.~(\ref{err-opp}) and~(\ref{fid-perf}) that the fidelity
of the phase gate with $\pi$~pulses is given for $\delta\vartheta\ll 1$ as
\be
\label{non100}
{\cal F}'=\left\l\frac{1}{2}[1+\cos(\delta\vartheta)]\right\r\approx
1-\frac{1}{4}\left\l(\delta\vartheta)^2\right\r\,,
\ee
where $\l\cdot\r$ denotes averaging as stated in Eq.(\ref{non0}).
Before we calculate the fidelity it is convenient to relabel 
the~quantity~$\theta$, originally defined in Eq.~(\ref{ion30}), giving
\be
\label{non101}
\theta
=\left(
\sqrt{\frac{\pi}{8}}\epsilon\omega\tau\xi_0^2
\right)
\frac{\xi^2}{\xi_0^2}
=\theta_0\frac{\xi^2}{\xi_0^2}\,,
\ee
where $\xi_0$ refers to the amplitude of the force at $x=0$ ($z=0$) and
$\xi=\xi_{\rm trav}(x)$
or $\xi=\xi_{\rm stan}(z)$, respectively. Carrying out the~averaging in
Eq.~(\ref{non100}) the fidelity of the phase gate with $\pi$~pulses
for the non-unitary pushing force up to $(a/d)^4$ is found to be
\be
\label{non102}
{\cal F}'
&=&
1-\frac{\theta_0^2}{4\xi_0^4}
\bigg\{
\left[\bar{\xi^4}-(\bar{\xi^2})^2\right]
\nonumber\\[1mm]
& &
\!\!
+\left(\frac{a}{d}\right)^2
\left[
2\left(\frac{12k_BT}{\hbar\omega}\right)
\left[\bar{\xi^4}-(\bar{\xi^2})^2\right]
+\frac{2}{\sqrt{2}}\left(\bar{\xi^6}-\bar{\xi^2}\bar{\xi^4}\right)
\right]
\nonumber\\[1mm]
& &
\!\!
+\left(\frac{a}{d}\right)^4
\left[
\left(\frac{12k_BT}{\hbar\omega}\right)^2
\left[2\bar{\xi^4}-(\bar{\xi^2})^2\right]
+\frac{2}{\sqrt{2}}\left(\frac{12k_BT}{\hbar\omega}\right)
\!\!\times
\right.
\nonumber\\[1mm]
& &
\hspace{14mm}
\left.
\times
\left(\bar{\xi^6}-\bar{\xi^2}\bar{\xi^4}\right)
+\frac{1}{2}\left[\bar{\xi^8}-\big(\bar{\xi^4}\big)^2\right]
\right]
\bigg\},
\ee
where
\be
\label{non103}
\bar{\xi^n}\equiv
\int\limits_{-\infty}^{+\infty}
dx\,\xi^n(x)\,P(x)\,.
\ee

To calculate the quantity $\bar{\xi^n}$ which appears in the expression for
the fidelity, we use Eq.~(\ref{non3}) for the case of the travelling wave.
Then, in the limit $r\ll 1$, where we introduce the coefficient
$r\equiv 2\sigma/w\approx 2a\sqrt{\l n\r}/w$, we find
\be
\label{non104}
\bar{\xi^n}\approx
\xi_0^n\left[1+r^2A^{(n)}+r^4B^{(n)}\right]\,.
\ee
We will consider also the term of the order of $r^4$ because it scales with
$(a/w)^4$ and it is comparable with or larger than $(a/d)^4$.
Finally, we need to evaluate the coefficients $A^{(n)}$ and $B^{(n)}$ for
$n=2,4,6,8$. They are
\bse
\label{non105}
\be
A^{(2)}&=&\frac{1}{R^2}+2R^2-5\,,\\
B^{(2)}&=&-\frac{3}{R^2}+\frac{39}{2}-14R^2+2R^4\,,
\nonumber\\[2mm]
A^{(4)}&=&\frac{6}{R^2}+8R^2-18\,,\\
B^{(4)}&=&\frac{3}{R^4}-\frac{84}{R^2}+246-176R^2+32R^4\,,
\nonumber\\[2mm]
A^{(6)}&=&\frac{15}{R^2}+18R^2-39\,,\\ 
B^{(6)}&=&\frac{45}{R^4}-\frac{495}{R^2}+\frac{2295}{2}-810R^2+162R^4\,,
\nonumber\\[2mm]
A^{(8)}&=&\frac{28}{R^2}+32R^2-68\,,\\ 
B^{(8)}&=&\frac{210}{R^4}-\frac{1680}{R^2}+3480-2432R^2+512R^4\,,
\nonumber
\ee
\ese
where we denoted $R\equiv 2x_0/w$. It is of the order of one ($R\sim 1$)
because $x_0$ refers to the position of the ion in the laser beam and hence
we can assume $x_0\sim w$.

By similar means, we can calculate $\bar{\xi^n}$ for the case of the standing
wave. Substituting Eq.~(\ref{non10}) into Eq.~(\ref{non103}) we get
\bse
\label{non106}
\be
\bar{\xi^2}
&=&
\frac{\xi_0^2}{2\sin^2(2kz_0)}
\left[
1-e^{-8(k\sigma)^2}\cos(4kz_0)
\right],\\[3mm]
\bar{\xi^4}
&=&
\frac{\xi_0^4}{8\sin^4(2kz_0)}
\left[
3-4e^{-8(k\sigma)^2}\cos(4kz_0)
\right.
\nonumber\\[1mm]
& &
\left.
+e^{-32(k\sigma)^2}\cos(8kz_0)
\right],\\[3mm]
\bar{\xi^6}
&=&
\frac{\xi_0^6}{32\sin^6(2kz_0)}
\left[
10-15e^{-8(k\sigma)^2}\cos(4kz_0)
\right.
\nonumber\\[1mm]
& &
\left.
+6e^{-32(k\sigma)^2}\cos(8kz_0)-e^{-72(k\sigma)^2}\cos(12kz_0)
\right],\nonumber\\\\
\bar{\xi^8}
&=&
\frac{\xi_0^8}{128\sin^8(2kx_0)}
\left[
35-56e^{-8(k\sigma)^2}\cos(4kz_0)
\right.
\nonumber\\[1mm]
& &
+28e^{-32(k\sigma)^2}\cos(8kz_0)-8e^{-72(k\sigma)^2}\cos(12kz_0)
\nonumber\\[1mm]
& &
\left.
+e^{-128(k\sigma)^2}\cos(16kz_0)
\right].
\ee
\ese
It is useful to write the exponents in Eq.~(\ref{non106}) in the form
\be
\label{non107}
(k\sigma)^2\approx 2\eta^2\l n\r\,,
\ee
where $\eta=ka/\sqrt{2}$ is so called {\it Lamb-Dicke parameter},
$k=2\pi/\lambda$ is the wavenumber of the laser light
and all other constants are defined in
Eq.~(\ref{non0.3}). Then, we distinguish two extreme cases:
(i)~Lamb-Dicke regime ($\eta^2\l n\r\ll 1$) and (ii) outside
the~Lamb-Dicke regime ($\eta^2\l n\r\gtrsim 1$).
The former case corresponds to the situation where the spatial extent of
the vibrational motion of the ion is much smaller than the wavelength of
the~laser light ($a\sqrt{\l n\r}\ll\lambda$). It means that the ion is
cold enough to be well
localized in the standing-wave and not to travel across its nodes and
antinodes. In other words, in the~Lamb-Dicke regime the ion does not
experience the~non-uniform profile of the standing wave. Then, we have
$e^{-(k\sigma)^2}\to 1$, giving $\bar{\xi^n}\to\xi_0^n$ and
the~fidelity~(\ref{non102}) reduces to Eq.~(\ref{ion36}) which
describes the case of the uniform pushing force.

Outside the Lamb-Dicke regime we have the situation where
$e^{-(k\sigma)^2}\to 0$ and it leaves us only with the first terms in
Eqs.~(\ref{non106}). In the expression for the fidelity~(\ref{non102}) 
the first term dominates, giving
\be
\label{non108}
{\cal F}'
\approx
1-\frac{\theta_0^2}{4\xi_0^2}\left[\bar{\xi^4}-(\bar{\xi^2})^2\right]
=
1-\frac{\theta_0^2}{32\sin^4(2kz_0)}.
\ee
The condition for the phase gate gives $2\theta_0\approx\pi$ and
we choose $kz_0=\pi/4$, then ${\cal F}'\approx 0.92$.
\vfill


\end{document}